\documentclass[prb,amssymb,superscriptaddress,showpacs,floatfix,a4paper,twocolumn]{revtex4}

\usepackage{graphicx}
\usepackage{bm}
\usepackage{amsmath}

\usepackage{color}

\newcommand{\zetaff}{\zeta_{\mathrm{FF}}}
\newcommand{\zetaeq}{\zeta_{\mathrm{EQ}}}

\newcommand{\zetarmff}{\zeta_{\mathrm{FF}}}
\newcommand{\zetarm}{\zeta}
\newcommand{\zetarmeq}{\zeta_{\mathrm{EQ}}}

\newcommand{\zrmff}{z_{\mathrm{FF}}}
\newcommand{\zrm}{z}

\newcommand{\zetarpff}{\zeta_{\mathrm{FF}}^{\mathrm{RP}}}
\newcommand{\zetarp}{\zeta^{\mathrm{RP}}}
\newcommand{\zetarpeq}{\zeta_{\mathrm{EQ}}^{\mathrm{RP}}}

\newcommand{\zetath}{\zeta_{\mathrm{TH}}}
\newcommand{\zetal}{\zeta_{\mathrm{L}}}
\newcommand{\zth}{z_{\mathrm{TH}}}

\begin{document}

\title{Random-Manifold to Random-Periodic Depinning of an Elastic Interface}

\author{S. Bustingorry}
\affiliation{CONICET, Centro At{\'{o}}mico Bariloche, 8400 San
Carlos de Bariloche, R\'{\i}o Negro, Argentina}
\author{A. B. Kolton}
\affiliation{CONICET, Centro At{\'{o}}mico Bariloche, 8400 San
Carlos de Bariloche, R\'{\i}o Negro, Argentina}
\author{T. Giamarchi}
\affiliation{DPMC-MaNEP, University of Geneva, 24 Quai Ernest Ansermet, 1211
Geneva 4, Switzerland}

\date{\today}

\begin{abstract}
We study numerically the depinning transition of driven elastic
interfaces in a random-periodic medium with localized
periodic-correlation peaks in the direction of motion. The
analysis of the moving interface geometry reveals the existence
of several characteristic lengths separating different
length-scale regimes of roughness. We determine the scaling
behavior of these lengths as a function of the velocity,
temperature, driving force, and transverse periodicity. A
dynamical roughness diagram is thus obtained which contains, at
small length scales, the critical and fast-flow regimes typical
of the random-manifold (or domain wall) depinning, and at large
length-scales, the critical and fast-flow regimes typical of
the random-periodic (or charge-density wave) depinning. From
the study of the equilibrium geometry we are also able to infer
the roughness diagram in the creep regime, extending the
depinning roughness diagram below threshold. Our results are
relevant for understanding the geometry at depinning of arrays
of elastically coupled thin manifolds in a disordered medium
such as driven particle chains or vortex-line planar arrays.
They also allow to properly control the effect of transverse
periodic boundary conditions in large-scale simulations of
driven disordered interfaces.
\end{abstract}


\maketitle

\section{Introduction}
\label{sec:introduction}

The dynamics of elastic manifolds in disordered media have been
widely studied in relation with the physical properties of many
systems.
Magnetic\cite{lemerle_domainwall_creep,bauer_deroughening_magnetic2,yamanouchi_creep_ferromagnetic_semiconductor2,metaxas_depinning_thermal_rounding}
or
ferroelectric\cite{paruch_ferro_roughness_dipolar,paruch_ferro_quench,kleemann_review_ferroelectrics}
domain walls, contact lines of liquid
menisci\cite{moulinet_distribution_width_contact_line2,ledoussal_contact_line}
fluid invasion fronts in porous
media,~\cite{Martys1991,Hecht2004} and
fractures~\cite{bouchaud_crack_propagation2,alava_review_cracks,bonamy_crackilng_fracture,ponson_depinning,ponson_fracture}
can be modelled as elastic interfaces or lines. Such lines
would be flat if they were not under the usually unavoidable
action of quenched disorder. Periodic systems like charge
density waves,~\cite{nattermann_cdw_review} vortex lattices in
type II
superconductors~\cite{blatter_vortex_review,giamarchi_vortex_review,du_aging_bragg_glass}
or Wigner crystals~\cite{giamarchi_electronic_crystals_review}
can be also modelled as elastic manifolds embedded in random
environments. This manifold is described by the displacements
around the perfect periodic lattice that would exist in the
absence of disorder. No matter how weak is the
disorder,~\cite{larkin_ovchinnikov_pinning} in all these
systems the competition between elasticity and disorder gives
rise to rough structures and complex collective pinning
phenomena with interesting universal features.

Of special interest is the response of this kind of systems to
an external uniform field, able to drive the elastic manifold
in a given direction. Concrete examples are applied magnetic
fields on magnetic domain walls, applied electrical fields on
ferroelectric domain walls, fluid pressure on contact lines,
tension on fractures, electrical currents on vortex lattices in
superconductors, and electrical fields on charge density waves
and Wigner crystals. Indeed, such a probe would be rather
trivial if not because of the  presence of quenched impurities:
Disorder breaks the translation symmetry (though not in a
statistical sense), making the otherwise uniform displacement
of the manifold a complicated process involving many degrees of
freedom. Whether the elastic bonds of the manifold break or
support the tearing produced by the disorder, the resulting
flow can be plastic or elastic respectively, and in both cases
a rich disorder-induced out of equilibrium phenomena can
emerge. To understand these phenomena it was shown to be more
convenient to start by restricting the study to the more
tractable elastic flow case.

Elastic {\it depinning} is one of the most prominent and better
understood examples of collective pinning dynamic
phenomena.~\cite{kardar_review_lines,Fisher_review_collective_transport}
At zero temperature the external field must overcome a finite
threshold $f_c$ in order to force the pinned system to acquire
a finite steady-state velocity $v$. Below the depinning
threshold a finite velocity is only possible at a finite
temperature by thermal activation, due to the presence of many
metastable states separated by energetic barriers. These
barriers tend to diverge when decreasing the drive in the
so-called {\it creep}
regime,~\cite{ioffe_creep,nattermann_creep} strongly impeding
the motion at low driving forces, and tend to vanish at $f_c$
giving place to a \textit{thermal
rounding}~\cite{middleton_CDW_thermal_exponent} of the
depinning transition. These collective transport phenomena are
experimentally relevant since a finite velocity in this kind of
systems correspond to physical quantities (magnetization, or
polarization for domain walls, voltage for superconductors,
current for CDW) that can be readily measured.

From the statistical physics point of view the most remarkable
feature of the far from equilibrium steady-state motion near
the depinning threshold $f_c$ at zero temperature is the
existence of a well defined non-trivial critical behavior. Just
above the threshold the motion is jerky, characterized by
forwardly moving avalanches of a typical size $\xi \sim
(f-f_c)^\nu$ and width $w \sim \xi^\zeta$ produced at a typical
rate $\tau \sim \xi^z$, yielding a mean velocity $v \sim
(f-f_c)^\beta$, with $\beta = \nu(\zeta-z)$. $\nu,\;z,\;\zeta$
are non-trivial characteristic exponents. These observations
led to the fruitful analogy of the depinning transition with
standard equilibrium critical phenomena, with $v$ playing the
role of the order parameter and $f$ the role of the control
parameter.~\cite{fisher_depinning_meanfield} This analogy
motivated an outburst of analytical and numerical work devoted
to determine the value of critical exponents for different
universality
classes,~\cite{kardar_review_lines,ledoussal_frg_twoloops,narayan_fisher_cdw,nattermann_stepanow_depinning,chauve_creep_short,chauve_creep_long,chen_marchetti,vandembroucq_thermal_rounding_extremal_model,nowak_thermal_rounding,roters_thermal_rounding1,roters_thermal_rounding2,rosso_dep_exponent,rosso_depinning_simulation,rosso_hartmann,duemmer2,bustingorry_thermal_rounding_epl}
and to develop powerful
analytical~\cite{middleton_theorem,fisher_functional_rg,balents_largeNexpansion}
and
numerical~\cite{rosso_depinning_simulation,kolton_short_time_exponents,rosso_correlator_RB_RF_depinning,kolton_depinning_zerot2,kolton_creep_exact_pathways}
methods to obtain them. From the numerical viewpoint such a
study requires a precise determination of the critical
threshold $f_c$.~\cite{rosso_depinning_simulation}

For standard equilibrium phase transitions the low-temperature
phase can be characterized by equilibrium correlation lengths
separating the critical-looking short length-scales from the
low-temperature fixed-point dominated large length-scales. For
the depinning transition it was shown that $\xi$ also admits an
analogous purely geometric interpretation as a crossover length
in the average steady-state roughness of the (`ordered') moving
$v>0$ phase. The length $\xi$ separates the regime of critical
roughness at short length scales (i.e. with a roughness
exponent of the critical configuration at $f_c$)  from the
fast-flow roughness observed at large length-scales (i.e. with
a roughness exponent identical to the strongly driven
interface, $f \gg
f_c$).~\cite{chauve_creep_long,duemmer2,kolton_creep_exact_pathways}
The steady-state geometry thus contains information of the
velocity and there is no need to observe the transient
correlated process of an avalanche. More recently, however, the
analysis of the low-temperature averaged steady-state geometry
has shown that no divergent steady-state correlation
length-scale exists approaching the threshold from below, thus
breaking the naive analogy with standard phase transitions,
where two divergent length-scales are expected above and below
the critical
point.~\cite{kolton_depinning_zerot2,kolton_creep_exact_pathways}

Elastic depinning universality classes were shown to depend on
the dimension of the embedding space $D$, the dimension $d$ of
the manifold or the number $N=D-d$ of displacement components
of the manifold, the nature of the elastic
interactions,~\cite{cule_1d_elastic_chain,cule_1d_elastic_chain_long,rosso_dep_exponent,rosso_depinning_simulation}
the anisotropy of the medium\cite{tang_anisotropic_depinning}
and the nature of microscopic disorder
correlations.\cite{chauve_creep_long,fedorenko_longrange_correlated_disorder}
Considering for simplicity the case of $d$-dimensional directed
manifolds with $N=1$ living in an isotropic uncorrelated
disordered medium it is convenient to distinguish between two
prominent groups, according to the correlations of the
effective pinning force $F_p(u,{\bf r})$. This pinning force
acts on the manifold displacement field $u({\bf r})$, which
measures the distance between the distorted and the perfectly
flat manifold at the labelling point ${\bf r}$. On one hand the
pinning force on interfaces such as domain walls or contact
lines in random potentials usually display short-range
correlations reflecting the fact that the interface sees a
completely different disorder after shifting it a distance
bigger than a certain characteristic finite width $r_f =
\max[w,r_0]$, where $w$ is the domain wall-width and $r_0$ the
assumed finite correlation length of the disorder potential. We
use Random-Manifold (RM) to denote this group, and we do not
make distinction between the Random-Bond and Random-Field type
of disorder since at depinning, unlike statics, they are known
to merge into a single
class.~\cite{chauve_creep_long,rosso_numerical_depinning_correlator}
Interfaces in periodic potentials or periodic condensates such
as charge density waves or periodic chains of elastically
coupled objects on the other hand display an effective pinning
force with periodic correlations with a period $M$ representing
the period of microscopic potential in the first case and the
lattice spacing in the second case. We use Random-Periodic (RP)
to denote this group.

For short-range correlated isotropic disorder, the $N=1$ RM and
RP classes have been traditionally studied, both numerically
and analytically, using two paradigmatic models of disorder.
While for modelling the large-scale dynamic behavior of a
non-periodic system it is enough to use any uncorrelated
potential with range $r_f$, for modelling the periodic system
the random-phase cosine potential have been traditionally
chosen, thus forcing $M \sim r_f$. Although this is a good
approximation for charge density waves (cf. Fukuyama-Lee-Rice
model~\cite{fukuyama_cdw_model,lee_cdw_model}), this kind of
modelling does not permit however to study the interesting
situation that can appear in different periodic systems for
which the periodicity is much larger than the short-range
correlation length of the disorder correlator, i.e. $r_f \ll
M$. Indeed, when the autocorrelation of the pinning force is
periodic and displays sharply localized peaks this physical
situation, mostly analyzed for two component ($N=2$)
displacement fields, was shown to be relevant for describing
the statics of Wigner
crystals~\cite{chitra_wigner_long,giamarchi_electronic_crystals_review}
or vortex
lattices,~\cite{giamarchi_vortex_short,giamarchi_vortex_long}
where the lattice spacing $a_0$ can be made much larger than
the vortex core size or coherence length $\xi$ (cyclotron
radius for Wigner crystals) by simply tuning an external
magnetic field. In these cases the length-scale separation is
responsible for the so-called RM regime of roughness. This
regime occurs at intermediate length-scales, before the system
asymptotically reaches the so-called Bragg-glass or RP regime.
Because the intermediate RM regime can span a wide range of
lengths,~\cite{kim_rm_regime,dolz_collective_pinning} it can
affect the statics and dynamics properties of this kind of
systems and thus can be experimentally observed. We can
therefore expect additional geometrical crossovers around the
depinning transition in these systems. From a numerical point
of view, the effect of a periodicity $M \gg r_f$ has been
already analyzed in the critical depinning force distribution
in Ref.~\onlinecite{bolech_critical_force_distribution}.

Here we present a study of the finite velocity dynamics of a
simple RP system which includes localized periodic correlation
peaks with controlled periodicity $M$, yielding an interesting
multiscale behavior around depinning. Our main result is a
geometrical dynamical roughness diagram which contains, at
small length scales, the critical and fast-flow regimes typical
of the RM (or ``magnetic domain wall'') depinning, and at large
length-scales, the critical and fast-flow regimes typical of
the RP (or ``charge-density wave'') depinning. We argue that
our results are qualitatively valid for the family of
one-component periodic systems with localized correlations
peaks, such as chains of elastically coupled thin interfaces.
We compare in particular a driven chain of interacting
particles in a one-dimensional disordered potential with an
elastic line in disordered potential with periodic correlations
at a larger scale. Our results are particularly relevant for
properly controlling and interpreting the effect of periodic
boundaries conditions in large-scale driven interface
simulations.

\subsection*{Outline of the paper}
\label{sec:organization}

The paper is organized as follows. In Sec.~\ref{sec:rpsystems}
we describe the general class of random periodic systems with
localized periodic correlation peaks for which we argue our
general results apply. Then, Sec.~\ref{sec:structure_factor}
presents the general properties of the structure factor, which
will be used to analyze the geometry of rough interfaces. The
main result of this work is the dynamical roughness diagram
presented in Sec.~\ref{sec:roughness_diagram_and_scaling} based
on scaling arguments. Section~\ref{sec:numerical_simulations}
gives the details of the performed numerical simulations that
will be presented in Sec.~\ref{sec:numerical_results} and give
support to the proposed dynamical roughness diagram. Then, in
Sec.~\ref{sec:discussion} we will present a discussion of the
RM-RP crossover, the relation between the elastic string and
particle chain models, the extensions of the roughness diagram
to the creep regime, and the implications of our results to
numerical simulations with periodic boundary conditions.
Finally, Sec.~\ref{sec:conclusions} presents the conclusions of
the present work.

\section{Random periodic systems with localized correlation peaks}
\label{sec:rpsystems}

\begin{figure}[!tbp]
\includegraphics[width=8cm,clip=true]{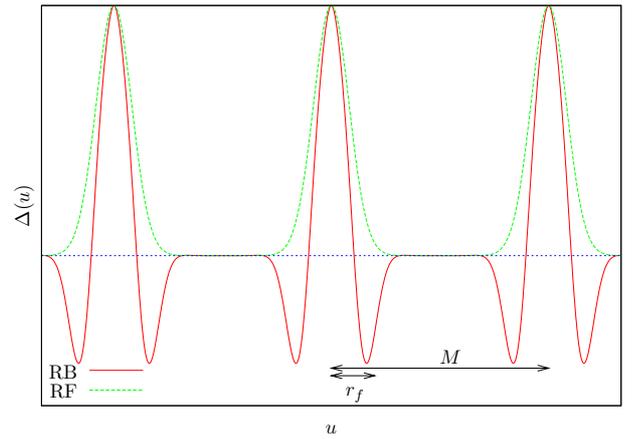}
\caption{(Color online) Schematic pinning force correlator for a random periodic system,
with periodicity $M$ and localized correlation peaks with range $r_f$.
RB and RF stand for random-bond and random-field correlations (see the text for details).}
\label{fig:scheme_correlator}
\end{figure}

As a model for a random periodic system with well separated
length scales $r_f$ and $M$ we focus our study on directed
elastic interfaces described by a one component displacement
field $u({\bf r},t)$ with internal dimension $d$, ${\bf r} \in
\Re^d$, which satisfy an overdamped equation of motion
\begin{equation}
\label{eq:interface_equation_of_motion}
\gamma \, \partial_t u({\bf r},t) = c \, \nabla^2 u({\bf r},t) + F_p(u,{\bf r})
+ f + \eta({\bf r},t),
\end{equation}
where $\gamma$ is the friction coefficient, $c$ the elastic
constant, and the uniform external force is given by $F$. The
thermal fluctuations satisfy
\begin{eqnarray}
\langle \eta({\bf r},t) \rangle &=& 0, \\
\langle \eta({\bf r},t) \eta({\bf r'},t') \rangle &=& 2 \gamma T \delta(t-t')
\delta({\bf r}-{\bf r'}),
\label{eq:thermal_noise_correlator}
\end{eqnarray}
and therefore the system asymptotically relax to the canonical
thermal equilibrium at temperature $T$ in the absence of the
driving force $f$. The pinning forces are characterized by
sample-to-sample fluctuations given by
\begin{equation}
\overline{F_p(u,{\bf r})F_p(u',{\bf r'})} = \Delta(u-u')\delta({\bf r}-{\bf
r'}).
\label{eq:pinning_force_correlator}
\end{equation}
In this paper we consider the case when the correlator function
$\Delta(u)$ is a periodic function with correlation peaks
localized in a range $r_f \ll M$ at the values $u=p M$ with $M$
the periodicity and $p$ any integer. We do not make a
distinction between the so-called random-bond where we must
enforce $\int du\;\Delta(u)=0$ and the random-field cases (see
e.g. Ref.~\onlinecite{chauve_creep_long}) since we are
interested in the depinning transition where these two
different static universality classes merge into a single one
(in the creep regime, for $f<f_c$ and $T>0$ the distinction
must be done however, since the static properties can affect
the intermediate length-scale
physics~\cite{chauve_creep_long,kolton_creep_exact_pathways}).
In Fig.~\ref{fig:scheme_correlator} we schematically represent
the shape of $\Delta(u)$.
A periodic pinning force with fluctuations given by
Eq.~\eqref{eq:pinning_force_correlator} arises naturally in
numerical simulations of interfaces in random environments,
when analyzing a system of transverse size $M$ with periodic
boundary conditions. We will exploit this fact to get most of
our numerical results.

Interestingly, as we will show later, we find that the
roughness scaling for elastic lines in a random-periodic two
dimensional potential with periodicity $M$ in the direction of
displacement also describe a chain of elastically coupled
particles in a one-dimensional non-periodic random potential,
with the lattice spacing given by $M$. We argue that this
connection is general, between  a $d$-dimensional elastic
manifold in a random-periodic medium and a periodic chain of
$(d-1)$-dimensional coupled manifolds. The dynamics of a
discrete chain of $d$-dimensional coupled manifolds can be
described by
\begin{eqnarray}
\label{eq:interface_chain_equation_motion}
\gamma \, \partial_t u_n({\bf r},t) = {\tilde c}[u_{n+1}({\bf r},t)+
u_{n-1}({\bf r},t) -2 u_{n}({\bf r},t)] \nonumber \\
+ c \, \nabla^2 u_n({\bf r},t) + G_p(nM + u_n,{\bf r}) + F + \eta_n({\bf r},t),
\end{eqnarray}
where $u_n$ describes the displacements of each manifold around
the perfect position $nM$ in the chain, $\tilde{c}$ is a
compression elastic constant, and $G(u,{\bf r})$ an
uncorrelated pinning force which is the same, independently of
$n$, with a \textit{non-periodic} short-range correlator of
range $r_f$. In this case, the correlations of the thermal
noise are given by $\langle \eta_n({\bf r},t)\eta_m({\bf r},t)
\rangle = 2\gamma T \delta_{nm}\delta({\bf r}-{\bf
r'})\delta(t-t')$. The connection between the physics of
Eq.~\eqref{eq:interface_equation_of_motion} for $d$-dimensional
manifolds and Eq.~\eqref{eq:interface_chain_equation_motion}
for $d-1$-dimensional manifolds is subtle since it involves a
non-trivial coarse-graining in the direction of the periodicity
which can produce extra terms in the equation of
motion.~\cite{giamarchi_vortex_short,giamarchi_vortex_long,cule_1d_elastic_chain_long}
This issue will be discussed in more detail later. It is
however plausible at this point that the resulting pinning
force would display, {\it if} distortions are locally smooth,
i.e. $|u_{n+1}-u_n| \ll M$, well developed periodic
correlations with a period $M$ in the direction of the chain
displacement with correlation peaks localized in a range $r_f
\ll M$, as the ones schematically shown in
Fig~\ref{fig:scheme_correlator}.

\section{Rough Geometry around Depinning}
\label{sec:structure_factor}

\begin{table}
 \caption{\label{table} Random-Manifold (RM) and Random-Periodic (RP) characteristic
 roughness exponents for the three reference states: equilibrium (EQ), depinning (dep), and fast-flow (FF).
 In the static equilibrium case only the Random-Bond (RB) class is quoted.}
 \begin{tabular}{|c|c|c|}
  \hline
  $d=1$ & RM & RP \\
  \hline
  EQ & $\zetarmeq = 2/3$(RB) & $\zetarpeq = 1/2$ \\
  dep & $\zetarm = 1.25$ & $\zetarp = 3/2$ \\
  FF & $\zetarmff = 1/2$ & $\zetarpff = 3/2$ \\
  \hline
 \end{tabular}
\end{table}

We focus our study on the geometrical observables that can be
defined for the sliding manifold. The structure factor $S_q$ is
a very convenient quantity to study the geometry of the
manifold at different length scales and to locate the different
crossovers.~\cite{bustingorry_thermal_rounding_epl,kolton_creep2,kolton_creep_exact_pathways,kolton_depinning_zerot2,kolton_flat_interface,kolton_short_time_exponents}
We define it as
\begin{equation}
S_q = \overline{\left \langle \left|\int e^{i q x} u({\bf r},t) d{\bf
r}\right|^2 \right \rangle}
\label{eq:structure_factor}
\end{equation}
where we have chosen the particular direction $x$ to be any
component of ${\bf r}$ if the interface is governed by
Eq.~\eqref{eq:interface_equation_of_motion} and the direction
of displacement (i.e. the direction of the chain) if it is
described by Eq.~\eqref{eq:interface_chain_equation_motion}.

The driven steady-state geometry at low temperatures is governed
by three reference states~\cite{kolton_depinning_zerot2,kolton_creep_exact_pathways}:
the  $f=0$ equilibrium state, the $f=f_c$ and $T=0$ depinning
critical state, and the fast-flow state $f\to \infty$.
The particularity of these states is that above a microscopic
length they have different self-affine geometries,
i.e. the structure factor behaves as
\begin{equation}
S_q \sim q^{-(d+2\zeta)},
\label{eq:self_afinne_scaling_Sofq}
\end{equation}
where the power-law behavior reflects the lack of a
characteristic length-scale in these states and $\zeta$ is the
characteristic roughness exponent.
The roughness exponents of the 
reference states are $\zetaeq$, $\zeta$ and $\zetaff$, respectively.
These exponents can take different values in
different universality classes. While $\zetaeq$ is different
for the RB, RF and RP universality classes, $\zeta$ and
$\zetaff$ remain the same for RB and RF classes and they change
for the RP class. Since we are particularly interested in
distinguishing the depinning and fast-flow roughness exponents
of the RP class and the RM class, we use a superindex ``RP'' in
all the exponent to indicate when the exponents belong to the
RP case, and omit the superindex for the RM class (see
Table~\ref{table}).

Furthermore, two important characteristic roughness exponents
are the Larkin exponent $\zetal$ and the thermal exponent
$\zetath$, which are in general expected to appear at very
small length-scales. The Larkin exponent is simply obtained by
doing a first order perturbation expansion in the disorder, thus
replacing it by a random uncorrelated force. It yields $\zetal
= (4-d)/2$ for lengths smaller than the Larkin length $l_c$,
above which the naive perturbation theory fails due to
metastability. The thermal roughness exponent $\zetath
=(2-d)/2$ is defined as the one that appears in absence of
disorder at finite temperature, and can be obtained exactly
from the Edwards-Wilkinson equation. Interestingly, we will
show later that both, $\zetal$ and $\zetath$ reappear at
\textit{large} length scales in the dynamics of a RP system
with localized correlation peaks: $\zetarmff=\zetath$,
$\zetarp=\zetarpff=\zetal$.

The steady-state geometry at small velocities can in general be
described by velocity and temperature dependent crossover
lengths separating different regimes of roughness. The
corresponding roughness exponents are however universal,
velocity and temperature independent, and coincide with one of
the aforementioned exponents. For velocities just above the RM
depinning transition we have
\begin{eqnarray}
S_q \sim
\left\{
\begin{array}{ll}
q^{-(d+2\zetarm)} & \text{for \,}q > 1/\xi \\
q^{-(d+2 \zetarmff)}& \text{for \,}q < 1/\xi,
\end{array}
\right.
\label{eq:roughness_crossover_rm_rmff}
\end{eqnarray}
allowing to define the characteristic length $\xi$. For small
velocities and vanishing temperatures $\xi$ can be identified
with a velocity dependent divergent correlation length~$\xi
\sim v^{-\nu/\beta}$. At $f \to f_c^+$ and zero temperature we
have $\xi \sim (f-f_c)^{-\nu}$,~\cite{duemmer2} with $v\sim
(f-f_c)^{\beta}$, and at $f=f_c$ and small temperatures we have
$\xi \sim T^{\psi
\nu/\beta}$,~\cite{bustingorry_thermal_rounding_epl} with
$\psi$ a thermal rounding exponent such that $v\sim T^\psi$.
Since in this case $S_q$ is governed by a single crossover
length $\xi$ we can write the scaling form
\begin{equation}
\label{eq:Sofq_with_xi}
S_q \sim \xi^{d+2\zetarm} \; s\left( q \xi \right),
\end{equation}
where the scaling function $s(x)$ behaves as $s(x) \sim
x^{-(d+2\zetarmff)}$ for $x \ll 1$ and $s(x) \sim
x^{-(d+2\zetarm)}$ for $x \gg 1$. By plotting $\xi$ vs $v$ we
can obtain a ``geometrical roughness diagram'' showing sectors
with different roughness exponents at different observation
length-scales $l$: $\zetarm$ for $l<\xi$ and $\zetarmff$ for
$l>\xi$. Physically, $\xi$ divides the small length-scales
which are dominated by the critical configuration, i.e. the
unique $v=0$ steady-state solution of the equation of motion
for $f=f_c$, from the large length-scales, which are governed
by an effective Edwards-Wilkinson equation with a velocity
dependent effective temperature dynamically induced by the
disorder.~\cite{chauve_creep_long} The physical origin of this
crossover is due to the fact that at small but finite velocity
the renormalized disorder becomes a weak perturbation at large
enough length scales, acting effectively as a thermal-like
noise in an Edwards-Wilkinson equation, with a short-range
correlation time of order $r_f/v$ and effective strength or
``temperature'' $\Delta(0)/v$.

In a random-periodic system with localized correlation peaks,
as the ones described in the previous section, the scaling of
Eq.~\eqref{eq:Sofq_with_xi} must be corrected to take into
account the existence of the additional characteristic distance
$M$. As we show in the next section, $M$ induces new
geometrical crossovers at depinning, separating the geometrical
roughness diagram in more than two sectors.

\section{Depinning roughness diagram and scaling arguments}
\label{sec:roughness_diagram_and_scaling}

In this section we summarize our most important physical
results about the steady-state geometry of driven
random-periodic systems with localized periodic correlation
peaks. We present the depinning roughness diagram and heuristic
scaling arguments describing the different crossovers. These
arguments are corroborated numerically and analytically in the
following sections.

\begin{figure}[!tbp]
\includegraphics[width=8.0cm,clip=true]{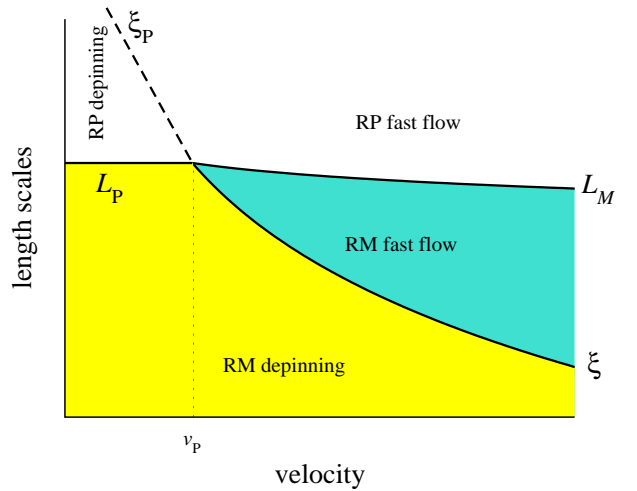}
\caption{(Color online) Schematic roughness diagram at the depinning transition
of a random-periodic system with periodicity $M$ and localized
periodic correlation peaks
as a function of its steady-state velocity $v$.
For $v < v_{\mathrm{P}}$, the geometry is of the RM (or domain-wall) class at
small length scales $l < L_{\mathrm{P}}$, while it is of the RP (or
charge-density-wave) class at large length
scales $l > L_{\mathrm{P}}$, where $L_{\mathrm{P}}\equiv L_P(M)$ but independent of th velocity $v$.
For $v > v_{\mathrm{P}}$, the geometry is of the RM class at small length scales
$l<\xi$, where $\xi\equiv \xi(v)$
is the RM depinning correlation length.
By further increasing the observation length-scale a crossover between
RM and RP fast-flow regimes of roughness
occurs at the length $L_M \equiv L_M(M,v)$.
Note that the large scale geometry is described by only one roughness
exponent since
fast-flow and depinning exponents coincide in the RP class, $\zetarp =
\zetarpff$.
The proposed scaling with the velocity $v$ and the periodicity $M$ of
the dynamical
crossover lengths $\xi(v)$, $L_{\mathrm{P}}(M)$, $L_M(v,M)$ (see text) is corroborated by
analyzing the structure factor obtained from numerical simulations.}
\label{fig:roughness_diagram}
\end{figure}
In Fig.~\ref{fig:roughness_diagram} we schematically show the
geometric roughness diagram we find, by analyzing the structure
factor, for a random periodic system with localized correlation
peaks. It presents three roughness sectors, characterized by
the roughness exponents of the RM depinning $\zetarm$, the RM
fast-flow $\zetarmff$ and the RP fast-flow $\zetarpff$.
Interestingly, unlike the RM case, for the RP system $\zetarp =
\zetarpff$, and therefore there is no signature, in the
steady-state structure factor, of the divergent length-scale
$\xi_{\mathrm{P}}$ expected for the depinning transition of a
pure RP system. We discuss this issue later. Below a
characteristic velocity $v_{\mathrm{P}}$ the system crosses over, at a
characteristic velocity-independent length $L_{\mathrm{P}}$, from a small
length-scale regime with a roughness exponent $\zetarm$,
corresponding to the geometry of the RM critical configuration,
towards a regime with an exponent $\zetarp$, corresponding to
the RP critical configuration. Above $v_{\mathrm{P}}$ there are two
crossovers, at the characteristic velocity-dependent
length-scales $\xi$ and $L_M$. The first crossover is from a
regime characterized by the RM critical depinning exponent
$\zetarm$ to a regime with the fast-flow RM exponent
$\zetarmff$. The second crossover, observed by further
increasing the observation length-scale, is from the RM
fast-flow regime to the RP fast-flow regime, the latter
characterized by the exponent $\zetarpff$ dominating the
largest length-scales.

\begin{figure}[!tbp]
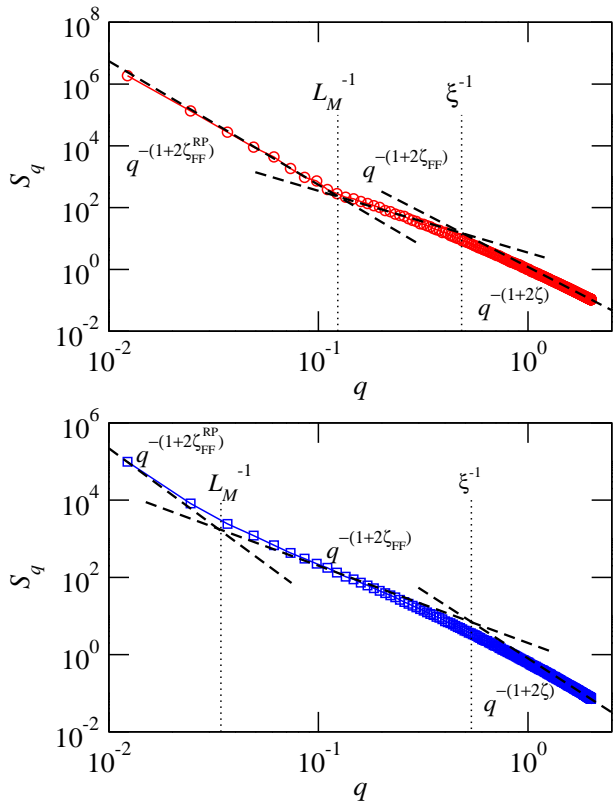

\includegraphics[width=8cm,clip=true]{sdeqdem-vfinita-fig3a}
\includegraphics[width=8cm,clip=true]{sdeqdem-vfinita-chain-fig3b}
\caption{(Color online) Typical structure factor for (a) a moving elastic string
in a periodic random medium with periodicity $M$, and (b) an elastic chain with lattice spacing $M$ moving
on a non-periodic disorder medium. The velocity of these systems is larger than their characteristic velocity
$v_{\mathrm{P}}$ (see Fig.~\ref{fig:roughness_diagram}). The two systems display a critical RM depinning roughness at small length scales
which crosses over to a RM fast-flow roughness at intermediate length scales, and then to a RP fast-flow roughness at the largest
length scales. These regimes are separated by two characteristic dynamical length scales: the correlation length $\xi$ and
the periodicity-induced length scale $L_M$.}
\label{fig:sofq_string_and_chain}
\end{figure}

The different length scales and roughness exponents shown in
Fig.~\ref{fig:roughness_diagram} can be obtained by analyzing
the structure factor. To illustrate how we obtain the roughness
diagram, in Fig.~\ref{fig:sofq_string_and_chain}(a) we show a
typical averaged structure factor for $v>v_{\mathrm{P}}$ for an interface
in a random periodic disorder medium with period $M$.
Increasing the observation length-scale (decreasing the wave
vector $q$) we see different crossovers between different
roughness regimes at the characteristic length-scales $\xi$ and
$L_M$. In Fig.~\ref{fig:sofq_string_and_chain}(b) we also show a
typical structure factor for an elastic chain with lattice
spacing $M$ moving in a one-dimensional disordered medium,
displaying identical regimes of roughness. This supports the
argued connection between the {\it geometrical} properties of
periodic chains of manifolds with internal dimensions $d$ in
$d+1$-dimensional pure random media and single interfaces of
internal dimension $d+1$ in $d+2$-dimensional random-periodic
media.

The geometric roughness diagram for a RP system with localized
correlation peaks is richer than for the RM system, which only
displays one critical characteristic length-scale $\xi$ above
$f_c$. This is due to the fact that $M$ is an extra
characteristic length in the problem, different from $r_f$,
unlike what occurs in CDW systems. Interestingly, the RP system
we study thus contains the RM depinning diagram, with its
$\zetarm$ and $\zetarmff$ roughness sectors, for velocities
larger than $v_{\mathrm{P}}$ and lengths below $L_M$. Indeed, we find $\xi
\sim v^{\nu/\beta}$ (independent of the periodicity $M$) which
coincides with the divergent length scale of the RM depinning,
being $\nu$ and $\beta$ RM critical exponents. We can thus say
that below $L_M$ for $v>v_{\mathrm{P}}$ and below $L_{\mathrm{P}}$ for $v<v_{\mathrm{P}}$
periodicity effects, both spatial and temporal, are not
important. In other words, short lengths scales decorrelate
rapidly, spatially and temporally, and can not sense the
periodic correlations of the pinning force.

To understand $L_{\mathrm{P}}$ we can make the simple scaling hypothesis
that periodicity effects start to be important when the average
transverse size or width of the critical manifold is of the
order of the periodicity $M$, $w \sim M$. Since at small
lengths $l$, the width grows as $w = r_f (l/l_c)^{\zetarm}$
(both for $v<v_{\mathrm{P}}$ and $v>v_{\mathrm{P}}$) we can compute the
characteristic length $L_{\mathrm{P}}$ by stating that $M\sim r_f
(L_{\mathrm{P}}/l_c)^{\zetarm}$. We thus obtain that,
\begin{equation}
L_{\mathrm{P}} \sim M^{1/\zetarm},
\label{eq:Lp_scaling}
\end{equation}
independent of the velocity. This describes well our data as a
function of $M,v$. Physically, $L_{\mathrm{P}}$ thus represents the length
at which the critical configuration starts to see the periodic
spatial correlations of the pinning force. Note that for the
CDW case we have $M \sim r_f$ and thus $L_{\mathrm{P}} \sim l_c$.
Therefore we would not observe this RM critical sector for a
CDW.

To estimate $L_M$ for $v>v_{\mathrm{P}}$ we must be more careful. Indeed,
the static argument given above of matching the width $w$ of
the sliding manifold with $M$ is incorrect in this case. This argument would
give $M\sim r_f (\xi/l_c)^{\zetarm} (L_M/\xi)^{\zetarmff}$ or
$L_M \sim \xi [(M/r_f)(l_c/\xi)^{\zetarm}]^{1/\zetarmff}$, and
by using $\xi \sim v^{-\nu/\beta}$ we finally get $L_M\sim
v^{-\nu(1-\zetarm/\zetarmff)/\beta}$. Since in general $\zetarm
> \zetarmff$ we get the incorrect result that $L_M$ grows with
the velocity, inconsistent with our data. The error in making
such an argument comes from the fact that for $v>v_{\mathrm{P}}$ the
roughness of the interface is determined, above $L_M$, by the
temporal correlations of the pinning force (when seen from the
moving interface). Indeed, despite the fact that $w < M$ and
that the renormalized disorder is already weak at $L_M$,
periodicity effects are relevant above a certain length beyond
which the manifold has not time to relax all its modes after
moving by a distance $M$. The steady-state geometry of the
moving system thus probes the periodic correlations of the
pinning force. We must thus compare the typical relaxation time
in the RM fast-flow regime $\tau(l) \sim \tau_c
(\xi/l_c)^{\zrm} (l/\xi)^{\zrmff}$, for $\xi< l \leq L_M$, with
the ``time of flight'' $\tau_M = M/v$, being $\zrmff \equiv
\zth = 2$ the dynamical exponent of the fast-flow RM class,
$\zrm$ the dynamical exponent of the critical RM regime, and
$\tau_c$ a microscopic time. If these times equate at $L_M$,
\begin{equation}
L_M \sim M^{1/\zrmff} \; v^{-\chi}.
\label{eq:LM_scaling_scaling}
\end{equation}
with
\begin{equation}
\chi=\frac{1}{\zrmff}-\frac{\nu}{\beta}\left(\frac{\zrm}{\zrmff}-1\right).
\label{eq:LM_scaling_scaling_exponent_chi}
\end{equation}
This result with $\chi>0$ describes well our numerical data, as
we show later.

Having $L_M$ and $L_{\mathrm{P}}$ we can now determine the characteristic 
velocity $v_{\mathrm{P}}$ of the roughness diagram, defined as $L_M(v_{\mathrm{P}})=L_{\mathrm{P}}$. We get,
\begin{equation}
 v_{\mathrm{P}} \sim M^{-1/\chi \zetarm},
\end{equation}
and therefore,
\begin{equation}
 L_M = L_{\mathrm{P}} \left( \frac{v_{\mathrm{P}}}{v}\right)^{\chi}.
\label{eq:LM_scaling_vs_LP}
\end{equation}

It is worth noting here that while $v_{\mathrm{P}}$ decreases, $L_M$ and
$L_{\mathrm{P}}$ increase with increasing $M$. This means that for large
enough $M$ the RM sector of the roughness diagram of
Fig.~\ref{fig:roughness_diagram} grows and in practice the
system behaves as a RM system. Conversely, for small $M$ the RP
sector grows and dominates the behavior at small velocities.

We also note that below $v_{\mathrm{P}}$ and above $L_{\mathrm{P}}$ we expect to
observe RP or CDW-like depinning, with a divergent correlation
length $\xi_{\mathrm{P}} \sim (f-f_c)^{-\nu^{\mathrm{RP}}}$.
However, unlike the RM case, the divergent length does not
manifest itself as a crossover between roughness regimes of
the structure factor, since $\zetarp \equiv \zetarpff$.
This is consistent with the fact
that the roughness exponent $\zetarp=\zetal=(4-d)/2$ for the RP
appears in FRG calculations from the generation of a
random-force in the renormalized pinning
correlator.~\cite{ledoussal_frg_twoloops} In other words, the
pinning forces acting on pieces of size $L_{\mathrm{P}}$ are essentially
uncorrelated, and the model thus effectively becomes the Larkin
model, with a roughness exponent $\zetal=(4-d)/2$. In this
respect, in Sec.~\ref{sec:numerical_results} we show that
$\zetarpff \equiv \zeta_L$ from numerical simulations.

The roughness diagram of Fig.~\ref{fig:roughness_diagram}
appears to be valid at small finite temperatures within the
``thermal rounding'' regime, as we find numerically. In this
regime the effect of the temperature translates into a finite
velocity $v\sim T^{\psi}$ at $f=f_c$ but does not affect the
large-scale roughness regimes. The depinning roughness diagram
of Fig.~\ref{fig:roughness_diagram} thus remains the same,
whether the velocity is originated by driving force, small
temperature or both.

In the following sections we describe our numerical simulation
method and results supporting the roughness diagram of
Fig.~\ref{fig:roughness_diagram} and the scaling for the
different crossover lines.

\section{Details of Numerical Simulations}
\label{sec:numerical_simulations}

We present here a detailed description of the numerical methods
we use to study the RP system with localized correlation peaks.
For simplicity we analyze low dimensional manifolds but our
results remain qualitatively the same for higher dimensions.
The $d=1$ case turns out to be on the other hand the most
stringent case for our general arguments.

We study the motion of an elastic string in a disordered
environment described by
Eq.~\eqref{eq:interface_equation_of_motion} in $d=1$ ($D=2$).
In order to numerically solve
Eq.~\eqref{eq:interface_equation_of_motion} for the elastic
string we discretize the $D=2$ embedding medium in the
longitudinal $z$-direction in $L$ segments of unit size,
keeping the transverse displacement field $u(z)$ as a
continuous variable in the $x$-direction. The discrete system
of equations read
\begin{eqnarray}
\label{eq:equation_motion_string}
\gamma \, \partial_t u(z,t) &=& c \,[u(z+1,t)+u(z-1,t)-2u(z,t)]
\nonumber \\
 &+& F_p(u,z)
+ f + \eta(z,t),
\end{eqnarray}
with $z$ an integer. Periodic boundary conditions of size $M$
(resp. $L$) are imposed for the transverse (resp. longitudinal)
system sizes. Besides avoiding boundary effects, this model
presents several advantages which have been exploited in
various ways for non-periodic
systems.~\cite{duemmer2,kolton_creep2,kolton_depinning_zerot2,kolton_short_time_exponents,bustingorry_thermal_rounding_epl}
The critical force and critical configuration for such finite
systems can be determined for each sample in polynomial time
with arbitrary precision by exploiting the Middleton
theorems.~\cite{rosso_dep_exponent} Moreover, the complete
sequence of metastable states below threshold, and in
particular the one dominating the creep motion at low
temperatures can be determined for each particular sample
exactly, by generalized, Middleton-like,
theorems.~\cite{kolton_depinning_zerot2,kolton_creep_exact_pathways}
This method thus allows for a well controlled analysis of key
properties such as the critical force statistics, the critical
exponents of the depinning transition, and the different
roughness crossovers.

To show the generality of our results we also study for
comparison the problem of an elastic chain by solving
Eq.~\eqref{eq:interface_chain_equation_motion} in $D=1$ and
$d=0$,
\begin{eqnarray}
\label{eq:equation_motion_chain}
\gamma \, \partial_t u_n(t) &=&
 \tilde{c} \,[u_{n+1}(t)+u_{n-1}(t)-2u_n(t)]
\nonumber \\
&+& G_p(nM + u_n)
+ f + \eta_n(t).
\end{eqnarray}
As
mentioned above, Fig.~\ref{fig:sofq_string_and_chain} shows
that the elastic string, described by
Eq.~\eqref{eq:equation_motion_string} and the elastic chain,
described by Eq.~\eqref{eq:equation_motion_chain} display the
same roughness crossovers around depinning. We argue that this
geometrical equivalence is general, between the $d$ dimensional
manifold in a $D=d+1$ disordered medium with period $M$ and the
periodic chain of $d-1$ dimensional elastically coupled
manifolds with lattice spacing $M$ in a $D=d$ dimensional
disorder medium. Therefore we study in details the case of the
elastic string and translate appropriately our results to both
kinds of systems in any dimension $d$. The resulting discrete
system of equations for the elastic chain in $D=1$,
Eq.~\eqref{eq:equation_motion_chain} are indeed similar to the
ones for the string in $D=2$, by identifying the discrete
values of $z$ for the particles of the string with the index
$n$ for the particles of the chain. The main difference between
the two systems are the pinning force correlations. While the
pinning force on the string is uncorrelated for different
values of the labelling variable $z$, the pinning force on the
chain is correlated for
different values of the labelling variable $n$, since in the
latter case the particles visit the same disorder as they move.
This difference can be better appreciated by remarking that the
equations for the elastic chain are equivalent (by interpreting
$n$ as $z$) to the ones of a tilted elastic string in a $D=2$
medium with columnar disorder, being $\theta= \tan^{-1} (M/L)$
the imposed tilting angle. The result of
Figs.~\ref{fig:sofq_string_and_chain} is thus non-trivial and
suggests that the roughness diagram of
Fig.~\ref{fig:roughness_diagram} is general for elastic system
with pinning forces displaying localized disorder correlation
peaks.

The equations of motion, Eqs.~\eqref{eq:equation_motion_string}
and \eqref{eq:equation_motion_chain}, are integrated using
Euler method with a time step $\delta t=0.01$. We set $\gamma
=1$, $c={\tilde c} =1$, $r_f=1$, and a disorder strength
$\Delta(0)=1$. A different choice of this microscopic
parameters does not qualitatively alter our results. The
continuous random potential for the string $V(u,z)=-\int
du\;F_p(u,z)$ is modelled by $L$ cubic splines passing through
$M$ regularly spaced uncorrelated Gaussian numbers points. For
the chain, the potential ${\tilde V}(u)=-\int du \;G_p(u)$ is
numerically generated with random spline passing through $L
\times M$ regularly spaced uncorrelated Gaussian numbers
points. Disorder average is done by averaging over different
realization of the gaussian random points. Using these disorder
potential models, when $M \gg r_f$ the corresponding pinning
forces display periodic correlations with localized peaks in a
range $r_f$.

\section{Numerical Results}
\label{sec:numerical_results}

In this section we show and discuss the numerical results for
the characteristics lengths, roughness exponents and
characteristic velocities, appearing in the geometrical
roughness diagram of Fig.~\ref{fig:roughness_diagram}. We
describe separately the different crossovers for $v>v_{\mathrm{P}}$ and
$v<v_{\mathrm{P}}$.

\subsection{Roughness Crossovers for $v>v_{\mathrm{P}}$}
\label{sec:roughness_crossovers_v_bt_vp}

\begin{figure}[!tbp]
\includegraphics[width=8cm,clip=true]{sdeqdem-vfinita-scal-MF-fig4}
\caption{(Color online) Scaling of the structure factor for
different $M$ values at  $T=0$ and $f = 1.2 f_c$. The longitudinal system size is $L = 512$. (a) Raw data. (b) Scaled data.}
\label{fig:Sofq_vs_M_f}
\end{figure}

\begin{figure}[!tbp]
\includegraphics[width=8cm,clip=true]{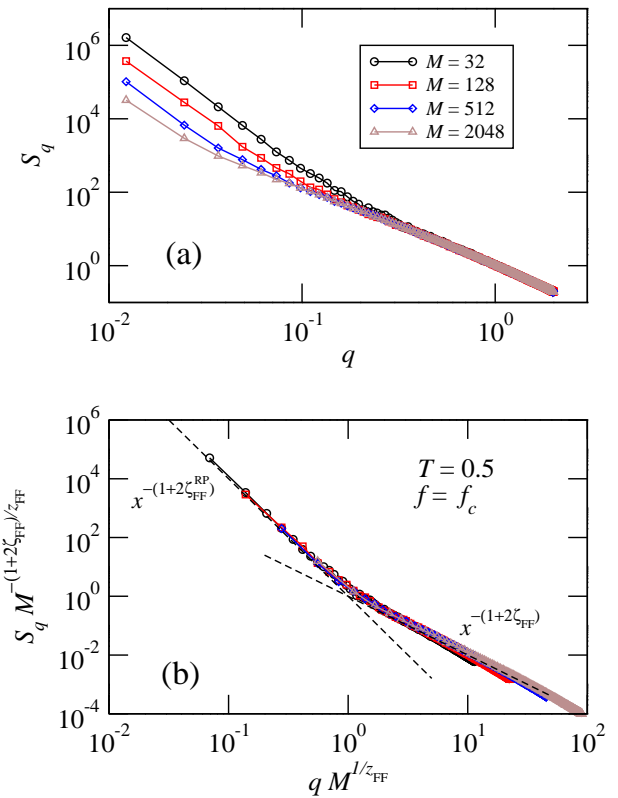}
\caption{(Color online) Scaling of the structure factor for
different $M$ values at $f=f_c$ and $T = 0.5$. The longitudinal system size is $L = 512$. (a) Raw data. (b) Scaled data.}
\label{fig:Sofq_vs_M_T}
\end{figure}

We start by discussing the two crossovers observed in
Figs.~\ref{fig:sofq_string_and_chain}(a) and
\ref{fig:sofq_string_and_chain}(b) for $v>v_{\mathrm{P}}$, at the
characteristic lengths $\xi$ and $L_M$ respectively. One
observes three roughness regimes for given values of $L$ and
$M$. They correspond to the three regimes observed in
Fig.~\ref{fig:roughness_diagram} for velocities above $v_{\mathrm{P}}$.
Increasing the length-scale (decreasing the wave vector $q$)
the local roughness exponent changes from $\zetarm \approx
1.25$, to $\zetarmff = 0.5$, and finally to $\zetarpff = \zetal
= 1.5$. The first two roughness exponents are characteristic of
the RM depinning, and we can identify the crossover length
$\xi$ with the divergent correlation length of the RM depinning
down to $v_{\mathrm{P}}$. The second crossover length $L_M$ is proper to
our system, separating the fast-flow regime of the RM class
from the one of the RP class.

For pure RM models the value $\zetarm \approx 1.25$ was
obtained numerically before by very different
methods,~\cite{rosso_roughness_MC,jensen_roughness_dep} and in
particular by exact algorithms.~\cite{rosso_hartmann,duemmer2}
Two loop renormalization group calculations for the RM class
are required to get values that are consistent with this
result.~\cite{ledoussal_frg_twoloops} The exponent $\zetarmff$
was obtained by numerical simulations,~\cite{duemmer2} and by
analytical arguments.~\cite{chauve_creep_long} As described
above the physical meaning of the appearance of $\zetarmff$ is
that at large length-scales the velocity $v$ becomes very
important and disorder effectively acts as an spatially
uncorrelated time-dependent perturbation with short-range
temporal correlations in a range $r_f/v$. The strength or
``effective temperature'' of this effective noise is thus
proportional to $\Delta(0)/v$.~\cite{chauve_creep_long}
Equation~\eqref{eq:interface_equation_of_motion} then
effectively becomes an Edwards-Wilkinson equation for which it
is straightforward to show that the steady-state roughness
exponent is $\zetath=(2-d)/2$ for $d\leq 2$, and $\zetath=0$
for $d>2$. For $d=1$ we have $\zetarmff = \zetath = 1/2$,
consistent with our numerical result for the present system.

\begin{figure}[!tbp]
\includegraphics[width=8cm,clip=true]{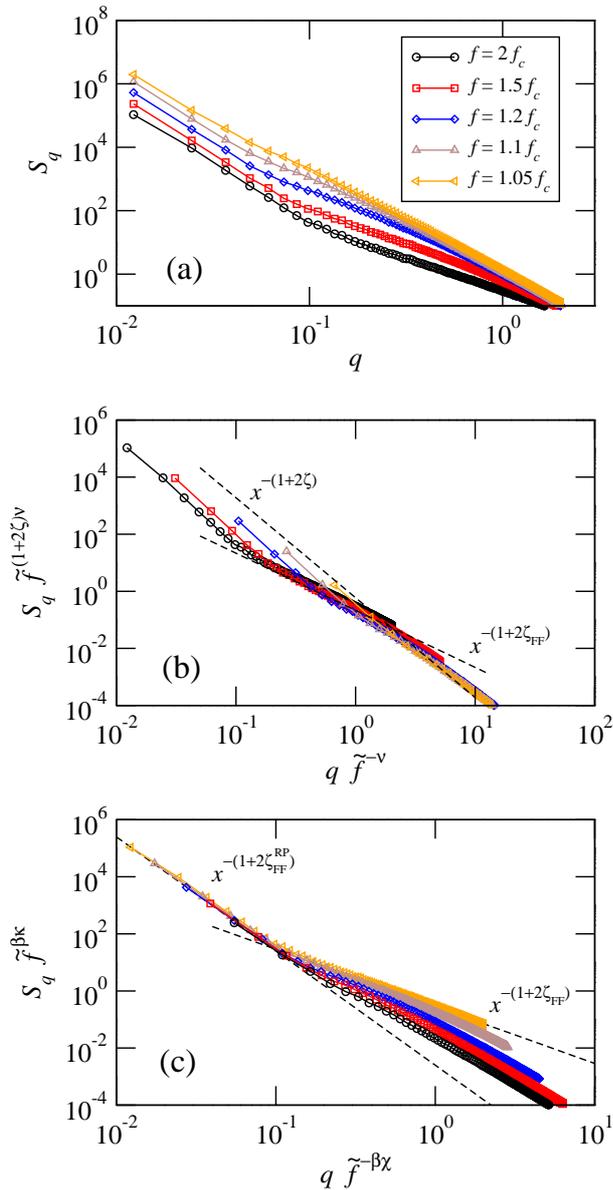}
\caption{(Color online) Scaling of the structure factor for different $f > f_c$ values at  $T=0$ and $M < L^{\zeta}$.
The longitudinal system size is $L = M = 512$. (a) Raw data, (b) scaled data around $\xi$, and (c) scaled data around $L_M$.}
\label{fig:Sofq_vs_f}
\end{figure}

\begin{figure}[!tbp]
\includegraphics[width=8cm,clip=true]{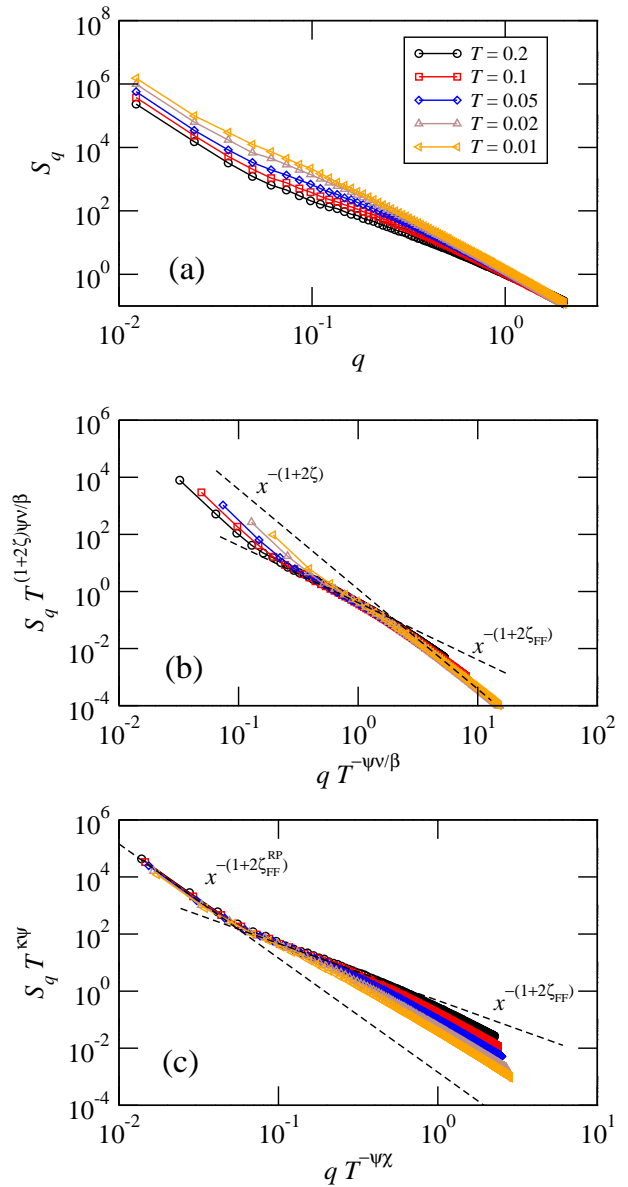}
\caption{(Color online) Scaling of the structure factor for different $T>0$ values at $f=f_c$ and $M < L^{\zeta}$.
The longitudinal system size is $L = M = 512$. (a) Raw data, (b) scaled data around $\xi$, and (c) scaled data around $L_M$.}
\label{fig:Sofq_vs_T}
\end{figure}

As described in Sec.~\ref{sec:roughness_diagram_and_scaling} the
crossover at $L_M$ for $v>v_{\mathrm{P}}$ occurs when the relaxation time
of the string in the RM fast-flow regime becomes of the order
of the time of flight $M/v$. The resulting scaling of $L_M$
with $M$ and $v$ involves several exponents of the RM class,
and hence it is a good test for the validity of our scaling
arguments. For fixed external force and temperature and
changing the transverse size $M$ only the crossover around
$L_M$ changes in the structure factor, as observed in
Figs.~\ref{fig:Sofq_vs_M_f}(a) and~\ref{fig:Sofq_vs_M_T}(a).
Thus, for the crossover at large length scales the structure
factor can be written, for $q \ll \xi^{-1}$ as,
\begin{equation}
 S_q L_M^{-(1+2\zetarmff)} = G(q L_M)
\label{eq:Sofq_with_LM}
\end{equation}
with $G(x) \sim x^{-(1+2\zetarpff)}$ for $x \ll 1$ and
$G(x)\sim x^{-(1+2 \zetarmff) }$ for $x \gg 1$. Using
Eq.~\eqref{eq:LM_scaling_scaling} for $L_M$, and since the
velocity is fixed by $f$ and $T$, we get the following scaling
formula,
\begin{equation}
S_q M^{-(1+2\zetarmff)/\zrmff} \sim  G(q \,M^{1/\zrmff}])
\label{eq:Sofq_scaling_LM}
\end{equation}
In Figs.~\ref{fig:Sofq_vs_M_f}(b) and~\ref{fig:Sofq_vs_M_T}(b)
we test this scaling prediction as a function of $M$ for a
fixed velocity by plotting $S_q M^{-(1+2\zetarmff)/\zrmff}$ vs
$q M^{1/\zrmff}$ for different values of $M$. In
Fig.~\ref{fig:Sofq_vs_M_f} the velocity is produced by a force
above threshold at zero temperature while in
Fig.~\ref{fig:Sofq_vs_M_T} the velocity is produced by a finite
but small temperature at $f=f_c$. In both cases we find that
the scaling form proposed collapses well the different curves
by using $\zrmff=2$ which corresponds to the dynamical exponent
of a RM model at large velocities (i.e. the dynamical exponent
of the Edwards-Wilkinson equation with the disorder-induced
Langevin-like noise). As expected, deviations from the good
collapse are observed only at large $q$, where the presence of
the extra characteristic length $\xi$ invalidates the simple
scaling of Eq.~\eqref{eq:Sofq_scaling_LM}. We also note that in
the non-scaled data in Figs.~\ref{fig:Sofq_vs_M_f}(a)
and~\ref{fig:Sofq_vs_M_T}(a) $S_q$ becomes $M$-independent for
$q \gg \xi^{-1}$. This is consistent with the fact that $\xi$
does not depend on $M$ but only on the velocity, $\xi \sim
v^{-\nu/\beta}$ near depinning, unlike $L_M$ which depends on
both, $v$ and $M$.

In order to study the velocity dependence of the
structure factor and its crossover lenghts
we have applied both different driving forces $f \gtrsim fc$
at $T=0$ and small temperatures for $f=f_c$.
In Figs.~\ref{fig:Sofq_vs_f}(a) and
~\ref{fig:Sofq_vs_T}(a) we show $S_q$ as a function
of the force and temperature, respectively.
The crossover around $\xi$ can be described, for $q \gg L_M^{-1}$,
with the scaling relation
\begin{equation}
S_q \sim v^{-(1+2\zeta) \nu/\beta} \; \tilde G\left[ q \, v^{-\nu/\beta} \right],
\end{equation}
with the function $\tilde G(x) \sim x^{-(1+2\zetaff)}$ for $x
\ll 1$ and $\tilde G(x) \sim x^{-(1+2\zeta)}$ for $x \gg 1$.
This scaling form depends on force and temperature only through
$v$. To get explicitly these dependencies we can use, for small
$v$ and $f \ge f_c$ that $v\sim \tilde f^{\beta}$ for $f
\gtrsim f_c$, with $\tilde f = (f-f_c)/f_c$ the reduced force,
and $v\sim T^{\psi}$ for $f = f_c$ and small
$T$.~\cite{bustingorry_thermal_rounding_epl}
Figures~\ref{fig:Sofq_vs_f}(b) and ~\ref{fig:Sofq_vs_T}(b) show
the respective scaling forms around $\xi$. However, this
scaling form is valid up to the scale $L_M$ where periodicity
effects are important and $\tilde G(x)$ is not longer
universal. The crossover of $\tilde G(x)$ to RP fast-flow can
be written as
\begin{equation}
 \tilde G(x) \sim x_M^{-(1+2\zetaff)} \overline G \left( \frac{x}{x_M} \right),
\end{equation}
where $x_M=v^{-\nu/\beta}/L_M$ and the new function $\overline
G(y) \sim y^{-(1+2\zetarpff)}$ for $y \ll 1$ and $\overline
G(y) \sim y^{-(1+2\zetaff)}$ for $y \gg 1$. Using
Eq.~\eqref{eq:LM_scaling_scaling} one can write,  for fixed $M$
and for $q \ll 1/\xi$ that the structure factor behaves as
\begin{equation}
S_q \sim v^{-\kappa} \overline G \left( q \, v^{-\chi} \right),
\end{equation}
with
\begin{equation}
 \kappa = 2(\zeta-\zetaff)\nu/\beta+(1+2\zetaff)\chi.
\end{equation}
This form describes the crossover between RM and RP fast-flow
regimes of the structure factor. Figures~\ref{fig:Sofq_vs_f}(c)
and ~\ref{fig:Sofq_vs_T}(c) show this scaling form when the
velocity is generated by finite drive at zero temperature or by
a small finite temperature at $f=f_c$.

Therefore, in Figs.~\ref{fig:Sofq_vs_f} and
~\ref{fig:Sofq_vs_T} we test the latter scaling prediction as a
function of $f$ for $f>f_c$ and $T=0$, and as a function of $T$
for $f=f_c$ by using the known values
$\beta=1/3$,~\cite{duemmer2,ledoussal_frg_twoloops}
$\psi=0.15$.~\cite{bustingorry_thermal_rounding_epl} Since
$S_q$ has two characteristic lengths $L_M$ and $\xi$, we show
separately the collapse around the two crossovers. In
Figs.~\ref{fig:Sofq_vs_f}(b) and ~\ref{fig:Sofq_vs_T}(b) we
show the collapse around $\xi$ and in
Figs.~\ref{fig:Sofq_vs_f}(c) and ~\ref{fig:Sofq_vs_T}(c) the
collapse around $L_M$, for the same set of curves $S_q(f,T)$ of
Figs.~\ref{fig:Sofq_vs_f}(a) and ~\ref{fig:Sofq_vs_T}(a),
respectively. The collapse obtained by using the known values
of $\psi$, $\beta$ $\zetarm$, $\zetarmff$, $\zrm$ and $\zrmff$
fully supports our interpretation of the two crossovers.

\subsection{Roughness crossovers for $v<v_{\mathrm{P}}$}
\label{sec:roughness_crossover_Lp}

\begin{figure}[!tbp]
\includegraphics[width=8cm,clip=true]{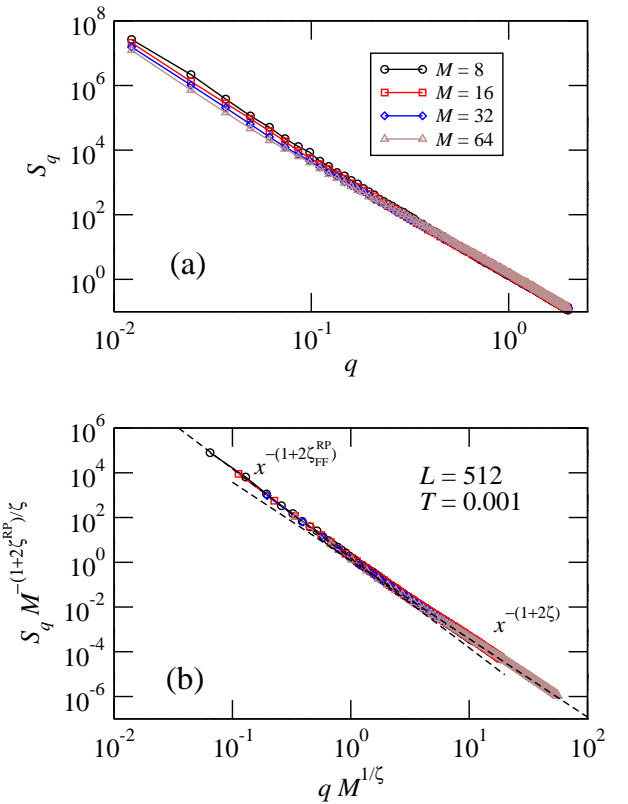}
\caption{(Color online) Scaling of the structure factor for different $M$ values in the $v < v_{\mathrm{P}}$ regime.
Data correspond to $T=0.001$ and $L = 512$. (a) Raw data. (b) Scaled data.}
\label{fig:Sofq_with_Lp}
\end{figure}

As shown in Fig.~\ref{fig:roughness_diagram}, at $v_{\mathrm{P}}$ we have
$L_M(v_{\mathrm{P}})=\xi(v_{\mathrm{P}})=L_{\mathrm{P}}$, where $L_{\mathrm{P}}$ is independent of the
velocity. The RM fast-flow regime thus disappears at $v_{\mathrm{P}}$.
This implies that periodicity effects are already generated by
static pinned configurations, and not dynamically as described
in the derivation of $L_M$. $L_{\mathrm{P}}$ is precisely the length at
which, for $v<v_{\mathrm{P}}$, the width of a typical RM critically pinned
configuration reaches $M$. Below $L_{\mathrm{P}}$, periodicity effects are
absent, and the typical critical configuration is not sensitive
to $M$. Just above $L_{\mathrm{P}}$, the critical configuration crosses
over to the RP class. To prove the existence of this crossover
we show in Fig.~\ref{fig:Sofq_with_Lp} that the structure
factor of a string with $L \gtrsim L_{\mathrm{P}}$ and $v<v_{\mathrm{P}}$ can be
written as
\begin{equation}
 S_q M^{-(1+2\zetarp)/\zetarm} = H(q M^{1/\zetarm})
\label{eq:Sofq_scaling_Lp}
\end{equation}
with $H(x) \sim x^{-(1+2\zetarp)}$ for $x \ll 1$ and $H(x)\sim
x^{-(1+2 \zetarm) }$ for $x \gg 1$. We have used $L\sim
M^{1/\zetarm}$ and the previously known values of $\zetarp$ and
$\zetarm$. The collapse of Fig.~\ref{fig:Sofq_with_Lp} thus
supports our interpretation of this crossover.

As discussed in Sec.~\ref{sec:roughness_diagram_and_scaling} and
shown in Fig.~\ref{fig:roughness_diagram}, we expect a
second crossover at a length $\xi_{\mathrm{P}}$ above $L_{\mathrm{P}}$, representing
the dynamical correlation of the RP depinning. The regime
between $L_{\mathrm{P}}$ and $\xi_{\mathrm{P}}$ thus represents the RP critical
regime, with the roughness exponent of the RP critical
configurations. As discussed in Sec.~\ref{sec:roughness_diagram_and_scaling}, 
$\zetarp=\zetarpff$, and due to this the RP depinning correlation length
$\xi_{\mathrm{P}}$ can not be detected by analyzing $S_q$. This
suggests that subtle geometrical measures are probably needed
to locate $\xi_{\mathrm{P}}$.~\cite{ledoussal_private}

\section{Discussion}
\label{sec:discussion}

In this section we discuss, by analytical arguments and
additional numerical simulations, the results obtained in the
previous sections. We first discuss, in
section~\ref{sec:rmtorf}, how to analytically calculate the
crossover from random-manifold to random-periodic fast flow
regimes of roughness at the length-scale $L_M$, which is absent
in the pure Random-Manifold or Random-Periodic depinning. The
numerical results of Sec.~\ref{sec:numerical_results} support
the depinning roughness diagram of
Fig.~\ref{fig:roughness_diagram} which is found to be the same
for interface pinning potentials with periodic correlations and
for periodic elastic systems such as chains. In
Sec.~\ref{sec:map} we discuss this interesting equivalence
between the geometry of periodic elastic systems in random
pinning potentials and interfaces in random-periodic pinning
potentials. We give analytical and numerical arguments. In
Sec.~\ref{sec:creepdiagram} we discuss how to extend the
roughness diagram of fig.~\ref{fig:roughness_diagram} to the
creep regime in the low temperature limit, based on the effects
of periodicity in the statics and depinning. Finally, in
Sec.~\ref{sec:periodic-boundary} we discuss the implications of
the roughness phase diagram for numerical simulations of
elastic strings with periodic boundary conditions and the
thermodynamic limit.

\subsection{Random-manifold to Random-periodic crossover in the fast flow regime}
\label{sec:rmtorf}

We show here how the crossover from random-manifold to
random-periodic fast-flow and their respective roughness
exponents can be analytically computed. We consider here a
simplified model of disorder perturbatively in the large
velocity and small temperature limit of
Eq.~\eqref{eq:equation_motion_string}. We will show this
approach yields correct  results for the velocity and
periodicity dependence of the crossover length, which can be
therefore identified with $L_M$ in
Fig.~\ref{fig:roughness_diagram}, when using the renormalized
friction, disorder strength and temperature at the length-scale
$\xi$.

At high velocities and small temperatures, we approximate the
pinning force $F_p(u,{\bf r})$ by $F_p(vt,{\bf r})$. In doing
so, we assume, \textit{a priori}, that the disorder-induced and
thermally-induced displacements are small in the regime we are
interested in. In this approximation the pinning force becomes
an effective thermal-like noise ${\tilde \eta}(t,{\bf r})
\equiv F_p(vt,{\bf r})$ with temporal correlations given by
\begin{equation}
 \overline{{\tilde \eta}(t,{\bf r}) {\tilde \eta}(t',{\bf r'})} = \Delta(v(t-t')) \delta({\bf r}-{\bf r'}).
\end{equation}
Since $\Delta(x)$ has a spatial range $r_f$, the range of
temporal correlations of $\tilde{\eta}$ is $r_f/v$ and its
effective temperature proportional to $\Delta(0)/\gamma v$.
Within this approximation, the equation of motion becomes
linear, and its solution in Fourier space is, for a particular
component $q={\bf q}.{\hat z}$ of the wave vector ${\bf q}$,
\begin{equation}
 u_q(t) =\gamma^{-1} \int_0^t dt' e^{-cq^2(t-t')/\gamma} \left[ f \delta_{q,0}
+ \tilde{\eta}_q(t) +\eta_q(t) \right].
\end{equation}
The instantaneous structure factor is thus given by
\begin{equation}
 S_q(t) = \overline{\langle |u_q(t)|^2 \rangle} = S^{\mathrm{EW}}_q(t) + S^{\mathrm{FF}}_q(t),
\end{equation}
where the first contribution is the Edwards-Wilkinson or purely thermal structure factor
\begin{equation}
 S^{\mathrm{EW}}_q(t)= \frac{T}{cq^2}\left( 1 - e^{-2cq^2t/\gamma} \right),
\end{equation}
and the second contribution comes from the disorder-induced noise ${\tilde \eta}$,
\begin{equation}
 S^{\mathrm{FF}}_q(t)= \gamma^{-1} \int_0^t \int_0^t dt_1\,dt_2 e^{-cq^2(2t-t_1-t_2)/\gamma}
\Delta[v(t_2-t_1)].
\end{equation}
To proceed we assume a particular periodic correlator
$\Delta(u)=\Delta(u+nM)$, with $n$ an integer. A simple choice
for $\Delta(u)$ having sharply localized peaks at $u=nM$ is
\begin{equation}
 \Delta(u) =\Delta_0 \sum_n \delta(u-nM),
\end{equation}
This kind of disorder arises physically from a random
distribution of identical point like pinning centers acting on
a very thin interface, in the limit $r_f \to 0$ with the
constraint $\int_0^M dx\; \Delta(x)=\Delta_0$ and $\Delta_0$ a
constant measuring the strength of the disorder. With such
disorder the fast-flow contribution to the structure factor can
be easily integrated to get
\begin{equation}
 S^{\mathrm{FF}}_q(t) = \frac{\Delta_0}{2 \gamma v c q^2 } \left( 1-e^{-2cq^2t/\gamma} \right)
\frac{1}{1-e^{-2cq^2M/\gamma v}}.
\end{equation}
Then, the total instantaneous structure factor can be expressed as
\begin{equation}
 \label{eq:sdeq-delta}
 S_q(t) = S^{\mathrm{EW}}_q(t)\left( 1 + \frac{\Delta_0}{2 \gamma vT} \frac{1}{1-e^{-2cq^2
M/\gamma v}}\right),
\end{equation}
Since we are interested in the steady-state we take the $t\to \infty$ limit to obtain the steady-state structure factor,
\begin{equation}
 S_q = \frac{1}{cq^2}\left( T + \frac{\Delta_0}{2\gamma v} \frac{1}{1-e^{-2 (q l_M)^2}}
\right),
\end{equation}
which presents the characteristic length $l_M \equiv \sqrt{M c/ \gamma v}$ or characteristic time $\tau_M \equiv M/v$.

For large length-scales such that $q \ll l_M^{-1}$ we have
\begin{equation}
 \label{eq:large_v_small_q_Sq}
 S_q \approx \frac{T}{c} q^{-2} + \frac{\Delta_0}{4 c M} q^{-4},
\end{equation}
where the velocity $v$ does not appear explicitly. At $T=0$,
$S_q \sim q^{-4}$, implying that the large scale roughness is
identical to one of the Larkin model $\zetal=(4-d)/2$ if $d\leq
4$ and $\zetal=0$ otherwise. This is consistent with our
finding $\zetarpff=\zetal=3/2$ for the $d=1$ case (string) in a
random periodic medium and also for the elastic chain. If $T>
0$ we could have a roughness crossover at $q_T =
\sqrt{\frac{\Delta_0}{4 T M}}$, where the two terms in
Eq.~\eqref{eq:large_v_small_q_Sq} become equal. However, for
this roughness crossover between $\zetath$ (with
$\zetath=(2-d)/2$ for $d \leq 2$ and $\zetath=0$ otherwise) to
$\zetal$ to be observable we must require $q_T \ll l_M^{-1}$ or
equivalently $T \gg \Delta_0 c/v$. Since $\Delta_0/v$ can be
identified with the shaking temperature $T_{sh}$ of
Ref.~\onlinecite{koshelev_dynamics} the condition for observing
this crossover in this regime reads $T_{sh}(v) \ll T$, meaning
that disorder-induced fluctuations must be much smaller than
thermal fluctuations.

If $q \gg l_M^{-1}$ on the other hand, we have
\begin{equation}
\label{eq:large_v_large_q_Sq}
 S_q \approx \frac{1}{cq^2}\left( T + \frac{\Delta_0}{2v} \right).
\end{equation}
which is equivalent to the Edwards-Wilkinson structure factor
at an effective temperature
$T_{\mathrm{eff}}(v)=T_{\mathrm{sh}}(v)+T$, yielding a
roughness exponent $\zetath$.

From (\ref{eq:large_v_large_q_Sq}) and
(\ref{eq:large_v_small_q_Sq}) we thus see that if the
temperature is small compared to
$T_{\mathrm{sh}}(v)=\Delta_0/2v$ there is a roughness crossover
at the length-scale $l_M$ from $\zetarmff=\zetath$ to
$\zetarpff=\zetal$ when increasing the observation
length-scale. This is in agreement with the roughness diagram
of Fig.~\ref{fig:roughness_diagram} for $v>v_{\mathrm{P}}$ and lengths
above $\xi$. By comparing $l_M$ and $L_M$ we see that both
quantities have the same $M$ dependence, since $\zrmff=2$. The
explicit velocity dependence, even if it is a power law in both
cases, is different. In this model we find $l_M \sim (\gamma
v)^{-1/2}$, instead of $L_M \sim v^{-1/2+(\nu/\beta)(z/2-1)}$
which describes well our data and was predicted in
Eq.~\eqref{eq:LM_scaling_scaling} by pure scaling arguments. We
also note in this respect that the structure factor at small
$q$ predicted by the model appears to be velocity independent
and temperature dependent, in contrast to what we predict by
scaling arguments and what the data presented in
Sec.~\ref{sec:numerical_results} supports. These differences
can be directly attributed to the incorrect use of the bare
friction constant $\gamma$, disorder strength $\Delta_0$ and
temperature $T$ in our perturbation theory, instead of using
their renormalized velocity-dependent values
$\tilde{\gamma}(v)$, $\tilde{\Delta}_0(v)$ and $\tilde{T}$ at
the length scale $\xi$. To prove this we first note that at the
length-scale $\xi$ we have $\tilde{\gamma}(v) v = (f-f_c) =
\xi^{-1/\nu}$ and therefore
$\tilde{\gamma}(v)=v^{-(\nu/\beta)(2-z)}$. Then, by replacing
$\gamma$ by $\tilde{\gamma}(v)$ we get the same velocity and
periodicity dependence for $l_M$ and $L_M$. This  justifies the
identification of the crossover predicted with the present
model with the one from the RM to the RP fast-flow regimes
observed in the simulations, and estimated in
Sec.~\ref{sec:roughness_diagram_and_scaling} by physical
arguments.

\subsection{Elastic String vs Elastic Chains}
\label{sec:map}

We discuss here the equivalence observed between the geometry
of an elastic line in a two-dimensional RP potential and the
one of a periodic chain in a one-dimensional non-periodic
random potential. We argue that this connection is general,
between D-dimensional thin interfaces transversely displacing
in random-periodic D+1 dimensional spaces and a periodic array
of (D-1) dimensions coupled interfaces in a D-dimensional
random medium. The connection between the two systems is
however not trivial, since there is no exact mapping between
these two systems. We describe first the case of the statics,
which can be discussed in terms of the replicated Hamiltonian
and complemented with additional transfer-matrix numerical
calculations, and then the dynamics.

\subsubsection{Statics}

\begin{figure}[!tbp]
\includegraphics[width=8cm,clip=true]{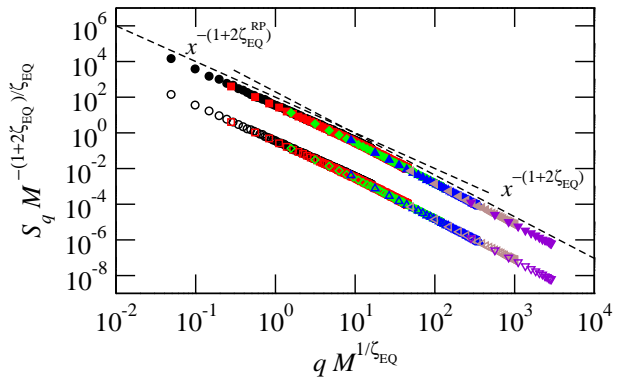}
\caption{(Color online) Ground state geometry of the elastic string (open symbols) and the one-dimensional chain (closed symbols)
problems as obtained with transfer matrix calculations. The data are represented in a scaled form and the closed symbols
are shifted upwards for clarity.}
\label{fig:GS-trasfer-matrix}
\end{figure}

We start by analyzing first the static problem, i.e. $F=0$ and
$T=0$, of the elastic line in a two dimensional random-periodic
disorder and then compare it with that of a one-dimensional
elastic chain over a random potential.

Let us consider an elastic line described by the univalued
function $u(z)$, where $u$ is a function in the transverse
direction, and $z$ gives the longitudinal direction. The
elastic contribution to the Hamiltonian is
\begin{equation}
 H_{\mathrm{e}} = \frac{c}{2} \int dz \, \left( \partial_z u \right)^2,
\end{equation}
while the disorder contribution in term of the line density $\rho(x,z)$ is
\begin{equation}
 H_{\mathrm{dis}} = \int dx\,dz\, V(x,z) \rho(x,z).
\end{equation}
Using the integral representation of the $\delta$ function one can write the density as
\begin{equation}
 \rho(x,z)=\delta(x-u(z))=\frac{1}{2 \pi}\int d\lambda\,e^{i \lambda(x-u(z))}.
\end{equation}

In order to consider the periodicity of the system in the
$x$-direction, the disorder potential can be written as a sum
over periodic images as
\begin{equation}
 V(x,z)=\sum_p \widetilde{V}(x-pM,z),
\end{equation}
where $\widetilde{V}(x,z)$ is the disorder potential defined in
the interval $x \in [-M/2,M/2]$. Using this periodic potential
one can write for the disorder Hamiltonian that (see the
Appendix)
\begin{equation}
 \label{eq:H-per}
 H_{\mathrm{dis}} =-\frac{\Delta}{2TM} \sum_{a,b} \sum_K \int dz\,
 e^{-i K(u_a(z)-u_b(z))},
\end{equation}
where $K=2 \pi n/M$. This last expression is strictly identical
to the one of a periodic system for which $K$ are the vectors
of the reciprocal lattice, as we show in the following.

In order to describe the one-dimensional periodic chain in a
one-dimensional disordered potential, let us consider an
elastic chain of average lattice space $a$, which corresponds
to average density $\rho_0 = 1/a$. One can imagine a chain
composed of masses and springs of constant length. Each 'mass'
has a finite internal width $\omega$ and the center of
consecutive masses are separated a distance $a$. The positions
of the particles are given by $x_j = x_j^0 + u_j = ja+u_j$,
where $x_j^0=ja$ is the nominal position in the unperturbed
lattice and $u_j$ is the displacement. In order to treat the
model one should go from the $u_j$ variables to a continuum
formulation. Through this relabeling process, the decomposition
of the density in this one dimensional problem leads to a
density field
\begin{eqnarray}
\label{eq:decomposition_density}
 \rho(x) &=& \sum_j \delta(x - x_j^0 - u_j) \nonumber \\
 &\cong& \rho_0 \left[ 1-\partial_x u(x) + \sum_{K \ne 0} e^{iK(x-u(x))}
\right],
\end{eqnarray}
where the last expression is valid for $\partial_x u \ll 1$,
and $K=2 \pi n/a$ are the reciprocal lattice vectors. The
continuum field
\begin{equation}
 u(x) =\int_0^{2 \pi/a} \frac{dq}{2 \pi} e^{i q x} \sum_j e^{i q x_j^0} u_j
\end{equation}
is valid for $x \gg a$. For details on the relabelling process
see Ref.~\onlinecite{giamarchi_vortex_long}, especially
Appendix A.

Considering the decomposition of the density and using a
replica formalism, the replicated disorder Hamiltonian can be
finally written as (see the Appendix)
\begin{widetext}
\begin{equation}
 H_{\mathrm{dis}} = -\frac{\Delta_0 \rho_0^2}{2T} \sum_{ab} \int dx\, \left[ \partial_x
u^a(x) \partial_x u^b(x) + \sum_{K \ne 0} e^{-iK (u^a(x)- u^b(x))}\right],
\end{equation}
\end{widetext}
which, with the exception of the $\partial_x u^a(x) \partial_x
u^b(x)$ term, is formally equivalent to Eq.~\eqref{eq:H-per}.
However, the original one dimensional chain in a one
dimensional disorder potential, as studied by Cule and
Hwa,~\cite{cule_1d_elastic_chain,cule_1d_elastic_chain_long}
contains also a term proportional to $\partial_x u(x)$. This
term is irrelevant at large length scales in dimensions $d>2$,
and gives a finite shift of the correlations function in $d \le
2$. Indeed it is commonly accepted that this term does not
change the roughness
exponent.~\cite{cule_1d_elastic_chain,cule_1d_elastic_chain_long}

Then, when comparing the periodic disorder case to the periodic
chain, the period of the disorder potential $M$ becomes the
average distance between neighboring particles. Thus, when the
fluctuation of the particles becomes of the order of the
average distance, the system enters a RP or CDW regime. Then,
in the static limit, structural fluctuations at large length scales are
given by the roughness exponent associated to the static CDW
problem, $\zetarpeq=1/2$. At finite $M>1$ a crossover appears
at a given length scale, corresponding to the scale for which
the disorder induced fluctuations become of the order of the
periodic box.

The geometrical equivalence argued above can be tested directly
by transfer matrix calculations for the ground-state of a
one-dimensional chain and the one of an elastic string. In
Fig.~\ref{fig:GS-trasfer-matrix} we compare the structure
factor of both systems, for different periodicities $M \gg
r_f$. We see almost no difference, meaning that the extra terms
in the replicated Hamiltonian for the chain are irrelevant. We
also find that the structure factor has a crossover at a
characteristic length $L_{\mathrm{P}}^0\equiv L_{\mathrm{P}}^0(M)$ between a regime
with the RM equilibrium roughness exponent $\zetarmeq=2/3$ at
small length scales, to a regime with the RP equilibrium roughness
exponent $\zetarpeq=1/2$ at large length scales. In
Fig.~\ref{fig:GS-trasfer-matrix} we show that the structure
factor is well described by the scaling formula
\begin{equation}
S_q \sim q^{-(1+2\zetarmeq)} \tilde H(q L_{\mathrm{P}}^0),
\end{equation}
with a crossover length $L_{\mathrm{P}}^0 \sim M^{1/\zetarmeq}$. The
function $\tilde H(x)$ is such that for small $x$ it behaves as
$x^{2(\zetarmeq-\zetarpeq)}$ The crossover at $L_{\mathrm{P}}^0$ can be
understood in the same terms as for the crossover length $L_{\mathrm{P}}$
at depinning. $L_{\mathrm{P}}^0$ is in this case the length at which the
width $w(L) \sim L^{\zetarmeq}$ of the interface at the
ground-state becomes of the order of the periodicity $w(L_{\mathrm{P}}^0)
\sim M$. As for depinning when $M \gg r_f$ the structure of the
system at equilibrium is identical to the RM one, and
crosses over to the RP one at large length-scales.

\subsubsection{Dynamics}

One can do a similar comparison for the dynamics of the elastic
line in the periodic disorder potential and the periodic chain
in disordered medium. As usual, we model the motion of an
elastic string in a disordered environment by means of the
overdamped equation of motion
Eq.~\eqref{eq:equation_motion_string}. The pinning force is
\begin{equation}
 F_p=-\frac{\delta H_{\mathrm{dis}}}{\delta u(z)}  =\frac{1}{L} \sum_K \widetilde{V}(x,z)
(iK) e^{i K (x-u(z))} e^{-K^2 \omega^2}.
\end{equation}
If now one use the Martin-Siggia-Rose formalism as in
Ref.~\onlinecite{ledoussal_mglass_long}, then it can be shown
that $F_p$ interacts only linearly with the operator $\hat
u_{zt}$. Thus one can replicate over different times in order
to obtain an averaged expression similarly to what was done in
the previous section, which results in a term proportional to
\begin{equation}
 -\frac{\Delta_F}{L} \sum_K \int dz\, dt_1 dt_2 K^2 e^{-iK(u(z,t_1)-u(z,t_2))}
e^{-2K^2\omega^2},
\end{equation}
where we used that
\begin{equation}
 \overline{F_p(x,z)F_p(x',z')}=\Delta{F}\delta(x-x')\delta(z-z').
\end{equation}
Again, this is equivalent, up to a factor proportional to
$\partial_x u(x)$, to the case of the one dimensional periodic
system.~\cite{ledoussal_mglass_long} The main difference is the
convective term, but it can be shown that it is irrelevant,
since an arbitrary shift $u(x) \to u(x) + f(x)$ leaves the
disorder term unchanged.~\cite{ledoussal_mglass_long}

\subsection{Roughness Diagram in the Creep Regime}
\label{sec:creepdiagram}

\begin{figure}[!tbp]
\includegraphics[width=8cm,clip=true]{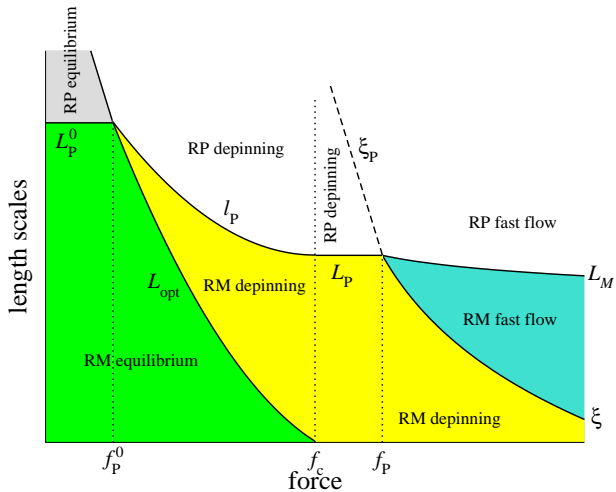}
\caption{(Color online) Depinning roughness diagram of
Fig.~\ref{fig:roughness_diagram} extended to the creep regime
$0<f<f_c$, at small temperatures.
The diagram displays roughness exponents of the RM and RP classes at
equilibrium, depinning and fast-flow.}
\label{fig:roughness_diagram_creep}
\end{figure}

The roughness diagram of Fig.~\ref{fig:roughness_diagram} for
depinning and the results of Fig.~\ref{fig:GS-trasfer-matrix}
can be combined to infer a roughness diagram as a function of
the driving force, including the expected crossovers in the
creep regime, $f<f_c$. As it was shown in
Ref.~\onlinecite{kolton_creep_exact_pathways} for pure RM (or
RF) systems, below the depinning threshold $f_c$ a
characteristic length $L_{\mathrm{opt}}\equiv
L_{\mathrm{opt}}(f)$ exists. This length grows with decreasing
$f$ separating the small length-scales described by the
equilibrium ($f=0$) roughness from the depinning ($f=f_c$)
roughness. Since $M \gg r_f$, we can expect to observe the same
behavior for the random-periodic system at intermediate and
small length-scales for which the RP systems behave as RM (or
RF) systems. Since at large length-scales the effect of
periodicity should always appear, it is plausible to connect
the crossover lengths $L_{\mathrm{P}}$ at $f\sim f_c$ and $L_{\mathrm{P}}^0$ at $f=0$
by a dynamic crossover length $l_{\mathrm{P}} \equiv l_{\mathrm{P}}(f,M)$ in the
creep regime. The width of the interface at $L_{\mathrm{P}}$ is thus given
by
\begin{equation}
M \sim (L_{\mathrm{P}}^0)^{\zetarmeq}
\end{equation}
if $f < f_{\mathrm{P}}^0$ and
\begin{equation}
M \sim L_{\mathrm{opt}}^{\zetarmeq} \left(\frac{l_{\mathrm{P}}}{L_{\mathrm{opt}}}\right)^{\zetarm}
\end{equation}
if $f_{\mathrm{P}}^0 < f < f_c$, with $f_{\mathrm{P}}^0$ a new characteristic force, defined by the condition $l_{\mathrm{P}}(f_{\mathrm{P}}^0,M)=L_{\mathrm{P}}^0(M)$. Therefore
\begin{equation}
l_{\mathrm{P}} \sim L_{\mathrm{opt}} \left( \frac{M}{L_{\mathrm{opt}}^{\zetarmeq}}\right)^{1/\zetarm},
\end{equation}
with $f_{\mathrm{P}}^0 < f < f_c$. In the diagram of
Fig.~\ref{fig:roughness_diagram_creep} we schematically show
the crossover length $l_{\mathrm{P}}$ separating the RM from the RP
depinning roughness in the creep regime. It shows several
sectors, including the equilibrium, depinning and fast-flow
geometries of both the RM and the RP (note however that the
fast-flow and depinning RP regimes have the same roughness
exponent).

Note that at variance with $L_M$, $l_{\mathrm{P}}$ can be obtained by a
static argument similar to the one used for $L_{\mathrm{P}}$ since the
velocity vanishes rapidly in the low temperature creep regime.
The system has thus enough time to relax all its (non-zero)
modes in the time of flight $\tau_M = M/v$. Note in this
respect that the largest length-scales obey an
Edwards-Wilkinson equation with correlated noise, and thus
their relaxation times are governed by a dynamic exponent,
contrarily to the zero-mode displacem

\subsection{Periodic boundary conditions and the thermodynamic limit}
\label{sec:periodic-boundary}

We present here a discussion concerning numerical simulations
and the proper thermodynamic limit. It is well known that when
using numerical simulations to describe interface depinning and
creep, a main shortcoming of the numerical method is coming
from the difficulty in taking the thermodynamic limit $M\to
\infty$ and $L\to \infty$. In order to perform a consistent
finite-size analysis one has to carefully specify how both $M$
and $L$ should tend to infinity, as different prescriptions for
the aspect ratio scaling lead to very different results.

It has been shown that the sample-to-sample fluctuations of the
critical force drastically change with the ratio
$M/L^{\zeta}$.~\cite{bolech_critical_force_distribution} For
very small $M$ compared to $L^\zeta$ periodicity effects are
important and the distribution of critical forces is Gaussian,
while at very high values of $M$ the critical force is
dominated by extreme values and its distribution becomes of the
Gumbel form. In the first case the mean critical force is
always bounded while in the latter grows logarithmically with
$M$ and is thus infinite in the thermodynamic limit. For the
aspect ratio scaling $M \sim L^{\zeta}$ it was shown that the
mean critical force is finite and well defined in the
thermodynamic limit where the system displays pure RM behavior.
In this last case, the critical force distribution is between
Gaussian and Gumbel. The aspect ratio scaling $M = k L^{\zeta}$
leaves however open the question of what is the optimal value
of $k$ for avoiding effects induced by the transverse boundary
conditions. This has motivated the use of a different method
for calculating steady-state properties of the same system, in
which the control parameter is not the force but the mean
velocity of the manifold by replacing the driving force by a
uniform spring with constant $m^2$, $f \to m^2 (vt-u({\bf
r}))$, pulling the whole manifold at a constant speed $v$ in a
transversely infinite medium. This model allows for more direct
comparisons with analytical
calculations,~\cite{ledoussal_avalanches,rosso_numerical_FRG,ledoussal_measuring_FRG,middleton_measuring_FRG,ledoussal_driven_particle,ledoussal_shocks,rosso_avalanche_FRG}
and has the advantage, compared with the force-controlled
model, that the characteristic length $L_m$ induced by the
parabolic moving potential is controlled only by the spring
constant $L_m \sim 1/m$, and not by the velocity-dependent
geometry of the manifold. When modelling a system for which the
spring has not a physical origin the correct scaling for the
spring constant is simply $L_m \sim L$ in this case and thus
very small values of $m$ are required in the large-scale limit.

If one wants to stick to the constant-force model the most
natural empirical choice for the aspect ratio scaling is $w(L)
\approx M$, where $w$ is the average width of the
manifold.~\cite{bolech_critical_force_distribution} The problem
with such prescription is that $w(L)$, being an integrated
quantity $w \sim \int dq\;S_q$, has a complicate dependence
with the mean velocity of the manifold through the
velocity-dependent correlations lengths separating different
regimes of roughness, and through the values of the different
roughness exponents. The geometrical roughness diagram of
Fig.~\ref{fig:roughness_diagram} shows clearly that this is
indeed the case, and gives at the same time an answer to this
problem. It shows that for a fixed value of $M$, the optimal
aspect ratio scalings are $L=L_{\mathrm{P}}(M)$ for $v<v_{\mathrm{P}}$, equivalent to
those proposed in
Ref.~\onlinecite{bolech_critical_force_distribution}, but a
different aspect-ratio scaling, $L=L_M(v,M)$, for $v>v_{\mathrm{P}}$.
Figure~\ref{fig:roughness_diagram} thus shows that using a
velocity independent prescription $L = L_{\mathrm{P}}(M)\sim M^{1/\zeta}$,
which works for the critical configuration, would always give
inconsistent results at all non-zero velocities in the
thermodynamic limit, since $v_{\mathrm{P}} \to 0$ when $M \to \infty$ and
then $L \gg L_M$ at a fixed $v$. Therefore, by increasing $M$
within this prescription the system would eventually display
periodicity induced effects at any finite $v$, inducing an
artificial crossover as a function of the velocity in
non-periodic systems. This crossover induced by periodicity is
on the other hand physically interesting for RP systems with
localized correlation peaks.

\section{Conclusions}
\label{sec:conclusions}

We have studied numerically the depinning transition of driven
elastic interfaces in a random-periodic medium with localized
periodic-correlation peaks in the direction of motion. We have
obtained a dynamical roughness diagram which contains, at small
length scales, the critical and fast-flow regimes typical of
the RM (or domain wall) depinning, and at large length-scales,
the critical and fast-flow regimes typical of the RP (or
charge-density wave) depinning. From the equilibrium behavior
of these kind of systems we have also inferred a richer
dynamical roughness diagram including the low temperature creep
regime which additionally includes roughness sectors
corresponding to the equilibrium geometry of the RP and RM
classes.

Our results are relevant for understanding the geometry at
depinning of periodic arrays of elastically coupled thin
manifolds in a disordered medium such as driven particle chains
or vortex-line planar arrays since these periodic systems
display localized periodic correlation peaks. In particular our
results are relevant for properly controlling the effect of
transverse periodic boundary conditions in large-scale
simulations of constant-force driven disordered interfaces.
From the roughness diagrams of
Figs.~\ref{fig:roughness_diagram} and
\ref{fig:roughness_diagram_creep} we see indeed that the aspect
ratio relation must be carefully chosen when taking the
thermodynamic limit, depending whether one wants to study the
large-scale behavior of a pure RM or a RP system.

We have also argued that there is a geometrical equivalence
between the d-dimensional periodic elastic system moving in
d-dimensions and the d-dimensional elastic interface moving in
a d+1 dimensional periodic medium, although the mapping between
these two systems is not exact. In this respect we note that
the $d=1$ case we have studied numerically is the most
stringent case, since it goes beyond the usual small slope
approximation used to develop the density in periodic
components, Eq.~\ref{eq:decomposition_density}. Indeed, since
the roughness exponent for chains is larger than one, the
average difference between the displacements of neighboring
particles grows with the system size, violating the small slope
approximation for large systems. Despite this fact, the results
still remain valid even for the one dimensional case. We thus
conclude that this equivalence is rather robust and should hold
for higher dimensional cases.

\begin{acknowledgments}
This work was supported in part by the Swiss National Science Foundation
under MaNEP and Division II. SB and ABK acknowledge financial support from ANPCyT Grant No. PICT2007886
and CONICET Grant No. PIP11220090100051.
\end{acknowledgments}

\appendix*

\section{Elastic String vs Elastic Chains}
\label{sec:ap-map}

In this Appendix we show how the disorder Hamiltonian for the
elastic string in periodic disorder and for the periodic chain
in one-dimensional disorder can be obtained.

\subsection{Two-dimensional periodic disorder}

The disorder contribution to the full Hamiltonian of the system is
\begin{equation}
 H_{\mathrm{dis}} = \int dx\,dz\, V(x,z) \rho(x,z).
\end{equation}
The line density $\rho(x,z)$ gives the position of the interface.

In order to consider the periodicity of the system in the $x$-direction, the disorder potential can be written as a sum over periodic images as
\begin{equation}
 V(x,z)=\sum_p \widetilde{V}(x-pM,z),
\end{equation}
where $\widetilde{V}(x,z)$ is the disorder potential defined in
the interval $x \in [-M/2,M/2]$. In terms of a traditional
uncorrelated Gaussian disorder $V_0(x,z)$, one can directly
define
\begin{equation}
 \widetilde{V}(x,z) = \Theta_M(x,z) V_0(x,z),
\end{equation}
using that
\begin{equation}
 \Theta_M(x,z) =
\left\{
\begin{array}{ll}
 1 & \text{if \,} x \in [-M/2,M/2], \\
 0 & \text{otherwise.} \\
\end{array}
\right.
\end{equation}
The disorder term thus becomes
\begin{eqnarray}
 H_{\mathrm{dis}}&=& \int dx\,dz\, V(x,z) \delta(x-u(z)) \nonumber \\
 &=& \int dx\,dz\, \Theta_M(x,z) V_0(x,z) \sum_p \delta(x-pM-u(z)). \nonumber \\
\end{eqnarray}

In order to obtain a disorder averaged Hamiltonian we use the replica-trick,~\cite{mezard_variational_replica,giamarchi_vortex_long}
\begin{equation}
 H_{\mathrm{dis}} = -\frac{1}{2T} \sum_{ab} \int dx\,dx'\, \overline{V(x)V(x')}
\rho_a(x)\rho_b(x').
\end{equation}
Then, using the integral representation of the $\delta$ function
\begin{equation}
 \rho(x,z)=\delta(x-u(z))=\frac{1}{2 \pi}\int d\lambda\,e^{i \lambda(x-u(z))},
\end{equation}
the replicated Hamiltonian reads~\cite{giamarchi_vortex_long}
\begin{eqnarray}
 H_{\mathrm{dis}} =&-&\frac{1}{2T} \sum_{a,b} \sum_{p_1,p_2} \int
dx_1\,dz_1\,dx_2\,dz_2\, \nonumber \\
 &\times&\Theta_M(x_1,z_1)\Theta_M(x_2,z_2) \nonumber \\
 &\times&\overline{V_0(x_1,z_1)V_0(x_2,z_2)} \nonumber \\
 &\times&\delta(x_1-p_1M-u_a(z_1)) \delta(x_2-p_2M-u_b(z_2)).\nonumber \\
\end{eqnarray}
Now, using for the disorder potential correlator that
\begin{equation}
 \overline{V_0(x_1,z_1)V_0(x_2,z_2)} = \Delta_0 \delta(x_1-x_2)\delta(z_1-z_2),
\end{equation}
and since $\Theta_M=0,1$ implies $\Theta_M^2=\Theta_M$, one has
\begin{eqnarray}
 H_{\mathrm{dis}} =&-&\frac{\Delta_0}{2T} \sum_{a,b} \sum_{p_1,p_2} \int dx\,dz\,
\Theta_M(x,z) \nonumber \\
 &\times& \delta(x-p_1M-u_a(z)) \delta(x-p_2M-u_b(z)).\nonumber \\
\end{eqnarray}
Now, we perform the sum over the localization function, resulting in
\begin{equation}
 \sum_{p_1} \delta (x-p_1M-u_a(z)) = \frac{1}{M} \sum_{K_1} e^{i K_1(x-u_a(z))},
\end{equation}
where $K_1 = 2 \pi n_1/M$, and we used that
\begin{equation}
 \sum_j e^{i q j M} = \frac{2 \pi}{M} \sum_K \delta(q-K).
\end{equation}
Thus, the disorder Hamiltonian can now be written as
\begin{eqnarray}
 H_{\mathrm{dis}} =&-&\frac{\Delta}{2TM^2} \sum_{a,b} \sum_{K_1,K_2} \int dx\,dz\,
\Theta_M(x,z) \nonumber \\
 &\times&e^{i K_1(x-u_a(z))} e^{i K_2(x-u_b(z))}.
\end{eqnarray}
Now, using that
\begin{equation}
 \int_{-M/2}^{M/2} dx\, e^{i(K_1-K_2)x} = M \delta_{K_1,K_2},
\end{equation}
performing the integral over $x$ and summing over $K_2$, one arrives at
\begin{equation}
 \label{eq:H-per-ap}
 H_{\mathrm{dis}} =-\frac{\Delta}{2TM} \sum_{a,b} \sum_K \int dz\,
 e^{-i K(u_a(z)-u_b(z))}.
\end{equation}

\subsection{One-dimensional periodic chain}

The positions of the particles are given by $x_j = x_j^0 + u_j
= ja+u_j$, where $x_j^0=ja$ is the nominal position in the
unperturbed lattice, $u_j$ is the displacement and $a$ is
the average lattice space. In order to treat the model one
should go from the $u_j$ variables to a continuum formulation.
Through this relabelling process, the decomposition of the
density in this one dimensional problem leads to a density
field
\begin{eqnarray}
 \rho(x) &=& \sum_j \delta(x - x_j^0 - u_j) \nonumber \\
 &\cong& \rho_0 \left[ 1-\partial_x u(x) + \sum_{K \ne 0} e^{iK(x-u(x))}
\right],
\end{eqnarray}
where the last expression is valid for $\partial_x u \ll 1$, $K=2 \pi n/a$, and the continuum field
\begin{equation}
 u(x) =\int_0^{2 \pi/a} \frac{dq}{2 \pi} e^{i q x} \sum_j e^{i q x_j^0} u_j
\end{equation}
is valid for $x \gg a$. This relabelling process is carefully
described in Ref.~\onlinecite{giamarchi_vortex_long},
especially Appendix A, and we refer the interested reader to
this work.

Considering the decomposition of the density, the Hamiltonian part corresponding to the uncorrelated Gaussian disorder $V_0(x)$ is
\begin{eqnarray}
 H_{\mathrm{dis}} &=& \int dx\, V_0(x)\rho(x) \nonumber \\
 &\cong& \rho_0 \int dx\, V_0(x) \left[ 1-\partial_x u(x) + \sum_{K \ne 0}
e^{iK(x-u(x))} \right]. \nonumber \\
\end{eqnarray}
When replicating the Hamiltonian one has
\begin{widetext}
 \begin{eqnarray}
  H_{\mathrm{dis}} &=& -\frac{\Delta_0 \rho_0^2}{2T} \sum_{ab} \int dx\, \left[
1-\partial_x u^a(x) - \partial_x u^b(x) + \sum_{K \ne 0} e^{iK(x-u^a(x))} +
\sum_{K' \ne 0} e^{iK'(x-u^b(x))} + \partial_x u^a(x) \partial_x u^b(x) \right.
\nonumber \\
  & & \quad - \left. \partial_x u^a(x) \sum_{K' \ne 0} e^{iK'(x-u^b(x))} -
\partial_x u^b(x) \sum_{K \ne 0} e^{iK(x-u^a(x))} + \sum_{K,K' \ne 0}
e^{-x(K+K')} e^{-i(K u^a(x)+K' u^b(x))}\right].
 \end{eqnarray}
In this last expression one should drop constant shift terms and rapidly oscillating terms. Then, setting $K=-K'$ in this last expression one has
\begin{equation}
 H_{\mathrm{dis}} = -\frac{\Delta_0 \rho_0^2}{2T} \sum_{ab} \int dx\, \left[ \partial_x
u^a(x) \partial_x u^b(x) + \sum_{K \ne 0} e^{-iK (u^a(x)- u^b(x))}\right].
\end{equation}
\end{widetext}

\bibliography{tfinita3}

\end{document}